\documentclass{aa}
\usepackage[utf8]{inputenc}
\usepackage[varg]{txfonts}
\usepackage{color}
\usepackage{amssymb}
\usepackage{amsmath}
\usepackage{array}
\usepackage[colorlinks=true,citecolor=blue,linkcolor=blue,urlcolor=blue]{hyperref}

\bibpunct{(}{)}{;}{a}{}{,}

\makeatletter
\renewcommand*\aa@pageof{, page \thepage{} of \pageref*{LastPage}}
\makeatother

\newcolumntype{L}[1]{>{\raggedright\let\newline\\\arraybackslash\hspace{0pt}}m{#1}}

\begin{document}

\title{The SRG/eROSITA all-sky survey: The morphologies of clusters of galaxies}
\subtitle{II. The intrinsic distributions of morphological parameters}
\titlerunning{eRASS1 cluster morphologies II}
\authorrunning{J.~S.~Sanders et~al.}

\author{
  J.~S. Sanders \inst{1}
  \and Y.~E.~Bahar \inst{1}
  \and E.~Bulbul \inst{1}
  \and N.~Clerc \inst{2}
  \and J.~Comparat \inst{1}
  \and M.~Kluge \inst{1}
  \and A.~Liu \inst{3,1}
  \and N.~Malavasi \inst{1}
  \and M.~E.~Ramos-Ceja \inst{1}
  \and T.~H.~Reiprich \inst{4} 
  \and F.~Balzer \inst{1}
  \and V.~Ghirardini \inst{5,1}
  \and F.~Pacaud \inst{4}
  \and X.~Zhang \inst{1}
}

\institute{
  Max-Planck-Institut f\"ur extraterrestrische Physik,
  Gießenbachstraße 1, 85748 Garching, Germany
  \and IRAP, CNRS, UPS, CNES, 14 Avenue Edouard Belin, 31400 Toulouse, France
  \and Institute for Frontiers in Astronomy and Astrophysics, Beijing Normal University, Beijing 102206, China
  \and Argelander-Institut für Astronomie (AIfA), Universität Bonn, Auf dem Hügel 71, 53121 Bonn, Germany
  \and INAF, Osservatorio di Astrofisica e Scienza dello Spazio, via Piero Gobetti 93/3, 40129 Bologna, Italy
  }

\date{Received ---, Accepted ---}

\abstract{
  X-ray selected surveys of clusters of galaxies have been reported to contain more regular cool core clusters compared to samples selected using the Sunyaev-Zel'dovich (SZ) effect.
  Morphology population studies on X-ray selected clusters will be biased without taking into account selection, as cool cores are more easily detected at low redshifts, but can be mistaken for point sources at high redshift.
  eROSITA, aboard Spectrum Roentgen Gamma (SRG), found over 12 thousand optically-identified clusters of galaxies in its first survey, eRASS1.
  Taking account of the selection function obtained from simulations, we obtain using a Bayesian framework the intrinsic distribution of morphological parameters, including the concentration, central density, cuspiness, ellipticity and slosh.
  We construct scaling relations for the parameters as a function of redshift and luminosity, and study their distribution within redshift or luminosity bins.
  We find that the concentration in a scaled aperture evolves positively with luminosity, similarly to the central scaled density, and negatively with redshift.
  When using a fixed aperture, its evolution with luminosity is lower, but also dependent on the choice of cluster centre.
  The mean cluster ellipticity does not significantly evolve with redshift or luminosity.
  eRASS1 clusters show indications of higher concentrations compared to SZ-selected objects, even after taking account the X-ray selection; this suggests that if our selection function model is correct SZ-selected clusters may also suffer from morphological selection effects.
  We compare different models for the parameter distribution in bins of redshift and luminosity.
  The distribution of concentration and ellipticity is generally consistent with a normal one, but other parameters such as the central density and cuspiness strongly favour more complex distributions.
  However, modelling of all clusters as a single population generally prefers non-normal distributions.
  }

\maketitle

\section{Introduction}
The morphology of galaxy clusters is a useful probe for understanding their dynamical state.
As clusters form from the hierarchical merger of galaxies, groups and clusters \citep[e.g.][]{Springel05}, cluster dynamical state is tied into their evolutionary history and is important for understanding how this hierarchical growth occurs \citep[e.g.][]{Evrard93,Wong12}.
Dynamical state can affect many aspects of clusters, including thermodynamic properties \citep[e.g.][]{Valdarnini21} and metallicities \citep[e.g.][]{Lovisari19}. It also affects cluster mass measurements, a key quantity for studying cluster populations \citep[e.g.][]{Pratt19}.
For example, dynamical state affects hydrostatic measurements, where the cluster is assumed to be in hydrostatic equilibrium \citep[e.g.][]{Biffi16}, cluster galaxy kinematics, which can be affected by mergers \citep[e.g.][]{Takizawa10}, and X-ray scaling relations, where luminosity in particular can be affected by the central gas density \citep[e.g.][]{EdgeStewart91},

The morphology of clusters can be characterised by a number of different parameters, including concentration \citep{Santos08}, central density \citep[e.g.][]{Lovisari17}, power ratios \citep{BuoteTsai95}, centroid shifts \citep[e.g.][]{BohringerPratt10}, ellipticity \citep[e.g.][]{Ghirardini22}, photon asymmetry \citep{Nurgaliev13}, Gini coefficient \citep{Lotz04} and cuspiness \citep{Vikhlinin07}.
These parameters are sensitive to different aspects of a clusters morphology.
For example, the concentration, central density and Gini coefficient are indicators of a cool core.
Power ratios, centroid shifts and photon asymmetry are sensitive to asymmetries in the 2D shape.

The extended ROentgen Survey with an Imaging Telescope Array (eROSITA; \citealt{Predehl21}) aboard the Spectrum Roentgen Gamma (SRG) observatory \citep{Sunyaev21} is an instrument designed to survey the X-ray sky.
The resulting catalogue of sources in the western Galactic hemisphere from its first sky survey was published in \cite{Merloni24}.
\cite{Bulbul24}, hereafter B24, created from these sources a catalogue of over twelve thousand optically-identified \citep{Kluge24} clusters of galaxies.
In \cite{Sanders25a}, hereafter Paper I, we studied the morphology of these clusters by measuring morphological parameters, including those listed above.
Where possible these parameters, including the concentration, central density, cuspiness and ellipticity, were measured using a forward-modelling technique, where the instrumental point spread function (PSF) and the sky background can be accounted for.
For measurements which are sensitive to the exact location of the cluster, we measure both an X-ray peak-centred quantity and one where the cluster position is included in the model.
Paper I also measures new forward-modelled parameters including slosh, how asymmetric a cluster is and multipole magnitudes, which are similar to power ratios.

Here, we build upon the work in Paper I to model the intrinsic distributions of morphological parameters, taking into account the selection effects present in the survey.
For example, we describe in Paper I how more concentrated clusters are more easily detected at low redshifts, as flatter objects can be mistaken as background variation, while the opposite is true at the highest redshifts, where peaked clusters can be misidentified as point sources.
To measure the true distribution we fit a model in which selection effects are properly included within a Bayesian framework.

Section \ref{sect:selfn_model} describes the selection function model.
The cluster subsamples and morphological parameters we analyse are discussed in Section \ref{sect:samp}.
The scaling relation model and results are given in Section \ref{sect:scaling_reln}.
We describe models using complex non-normal distributions in Section \ref{sect:complex_dist}.
Section \ref{sect:discuss} discusses our results and our conclusions are given in Section \ref{sect:concl}.

In this paper $\log$ refers to $\log_{10}$, while $\ln$ refers to $\log_\mathrm{e}$.
Uncertainties are at the $1\sigma$ confidence level unless otherwise specified.
We assume a cosmology with $H_0=70$~km~s$^{-1}$~Mpc$^{-1}$, $\Omega_\mathrm{m}=0.3$ and $\Omega_\Lambda=0.7$.
When computing the halo mass function, we assume $\sigma_8 = 0.8159$.

\section{Modelling of selection functions}
\label{sect:selfn_model}

Studying the intrinsic properties of the distribution of a parameter requires a selection function accounting for that parameter.
The standard selection functions for eROSITA clusters \citep{Clerc24} marginalise over the distribution of morphology included in the simulations of \cite{Comparat20}, including variation in ellipticity and realistic input profiles.
The selection function is also computed including a morphological parameter $EM_0$, which is the average emission measure within $0.025R_{500}$ and related to the central density.
However, as we investigated different morphological parameters and these parameters can affect detection in different ways and are not trivially converted to $EM_0$, it was necessary to make further simulations of clusters, where these parameters were measured, and build models of the selection function from the results.

The simulations described in Paper I did not exactly match the detection procedure used in the eROSITA catalogue because the \texttt{ermldet} detection software \citep{Brunner22} was used in imaging mode, rather than photon mode.
We improved our simulations to instead produce event lists so that the detection software could use photon mode, as tests showed differences in the obtained $\mathcal{L}_\mathrm{ext}$ values, affecting the cuts in selection.
The simulations are described in Appendix \ref{appen:sims}, and the selection function model in Appendix \ref{appen:seln}.

\begin{table}
  \caption{Subsamples from the first eROSITA all sky survey (eRASS1).}
  \centering
  \begin{tabular}{lr}
    \hline
    Subsample & Number \\ \hline
    All clusters & $12075$ \\
    $>50$ counts, cosmology & $2789$ \\
    $>100$ counts, cosmology & $541$ \\
    $\mathcal{L}_\mathrm{ext}>6$, $\mathcal{L}_\mathrm{det}>40$, cosmology & $2484$ \\
    $L_1$ ($41.1 \le L_\mathrm{X} < 43.3$, $>50$ cts., cosmo.) & $300$ \\
    $L_2$ ($43.3 \le L_\mathrm{X} < 43.7$, $>50$ cts., cosmo.) & $607$ \\
    $L_3$ ($43.7 \le L_\mathrm{X} < 44.0$, $>50$ cts., cosmo.) & $642$ \\
    $L_4$ ($44.0 \le L_\mathrm{X} < 44.3$, $>50$ cts., cosmo.) & $580$ \\
    $L_5$ ($43.3 \le L_\mathrm{X} < 45.6$, $>50$ cts., cosmo.) & $658$ \\
    $z_1$ ($0.0 \le z < 0.1$, $>50$ cts., cosmo.) & $56$ \\
    $z_2$ ($0.1 \le z < 0.2$, $>50$ cts., cosmo.) & $1000$ \\
    $z_3$ ($0.2 \le z < 0.3$, $>50$ cts., cosmo.) & $791$ \\
    $z_4$ ($0.3 \le z < 0.4$, $>50$ cts., cosmo.) & $489$ \\
    $z_5$ ($0.4 \le z < 1.5$, $>50$ cts., cosmo.) & $451$ \\
    \hline
  \end{tabular}
  \tablefoot{
    The subsamples marked cosmology were additionally selected to be part of the cosmology subsample from B24.
    The count selection is in the 0.2-2.3 keV band with an 800 kpc radius aperture.
  }
  \label{tab:subsamples}
\end{table}

\section{Fitted subsamples and parameters}
\label{sect:samp}
We took our input cluster sample from the eROSITA cosmology sample (B24), which already applies a cut to the cluster extension likelihood, $\mathcal{L}_\mathrm{ext}>6$, and redshift, resulting in a purer set of objects.
This likelihood was measured for each object with the detection likelihood ($\mathcal{L}_\mathrm{det}$) by \texttt{ermldet}.
From this sample we chose a brighter subsample, to avoid sources with large uncertainties on their morphological parameters and to avoid bias caused by preferential detection of steeply peaked cluster with high concentrations (Paper I).
We note, however, that we used the redshift \texttt{BEST\_Z} from the catalogue rather than the photometric redshift, unlike done for the cosmology analysis.
Other information taken from the catalogue are the cluster initial fit positions, redshifts, $L_{500}$, and $R_{500}$.

Rather than use $\mathcal{L}_\mathrm{det}$ to select brighter objects, we instead cut using the number of counts in an 800 kpc aperture, which is less affected by cluster concentration (See section 6.2 and figures 10 and 11 in Paper I).
At our 50 count cut, 68\% of the complete sample have $\mathcal{L}_\mathrm{det} = 32_{-16}^{+37}$, well above the minimum value of 5, while the corresponding range of $\mathcal{L}_\mathrm{ext} = 10.8_{-5.5}^{+6.3}$, which is less significantly above the threshold of 6 for the cosmology sample.
Taking the cosmology subsample and a minimum of 50 counts, so as not to be strongly affected by the standard initial $\mathcal{L}_\mathrm{det}$ threshold, results in a subsample of 2789 clusters (Table \ref{tab:subsamples}).
We also made a second subsample for consistency checks with a cut of 100 counts.

\begin{table}
  \caption{Summary of parameters.}
  \centering
  \begin{tabular}{lL{5.8cm}}
    \hline
    Parameter & Description \\ \hline
    $c_{500}$, $c_{500}^*$ & log concentration using apertures of $0.1R_{500}$ and $R_{500}$ \\
    $c_{80-800}$, $c_{80-800}^*$ & log concentration using apertures of 80 and 800~kpc \\
    $n_{\mathrm{s},0}$, $n_{\mathrm{s},0}^*$ & log gas density at $0.02R_{500}$ relative to the critical density \\
    $n_{50}$, $n_{50}^*$ & log electron density at a radius of 50~kpc \\
    $\alpha$, $\alpha^*$ & Cuspiness, or density slope, at $0.04R_{500}$ \\
    $\alpha_{50}$, $\alpha_{50}^*$ & Cuspiness, or density slope, at 50~kpc \\
    $\epsilon$ & Ellipticity, the ratio of minor to major axis (0-1) \\
    $H$ & Slosh (0-1; see Paper I) \\
    $M_{1}$ to $M_{4}$ & Multipole magnitudes (0-1; see Paper I) \\
    \hline
  \end{tabular}
  \tablefoot{
    Parameters with a ${}^*$ are measured with the cluster position fixed at the peak X-ray position rather than fitted for.
  }
  \label{tab:params}
\end{table}

In Paper I we measured a large number of parameters for each cluster.
We also described the various biases in measuring the parameters and showed that there are large systematic errors for all the non-forward modelled parameters, which vary with redshift and luminosity.
Rather than examine the intrinsic distributions for all these parameters, we restricted ourselves to the forward-modelled parameters listed in Table \ref{tab:params}.
For reasons of space we do not show the results for $M_1$-$M_4$ in all cases.
The forward modelled parameters are obtained through \texttt{MBProj2D} Markov chain Monte Carlo (MCMC) analyses (B24; Paper I), which allowed us to obtain the covariance between parameter values and luminosity.
The exception to these are the parameters $\epsilon$, $H$ and $M_1-M_4$, where we took the luminosity and morphology chains from separate runs and assume them independent.
This was done because a luminosity within a particular radius cannot easily be extracted using \texttt{MBProj2D} if the cluster is not circular.
At the minimum 50 count threshold, median uncertainties are around $0.23$ on the concentrations. They are $0.77$, $0.55$, $0.25$, and $0.17$ for $n_{\mathrm{s},0}$, $n_{\mathrm{s},0}^*$, $n_{50}$, and $n_{50}^*$, respectively. For $\alpha$, $\alpha^*$, $\alpha_{50}$, and $\alpha_{50}^*$, they are $1.1$, $0.9$, $0.9$, and $0.7$, respectively. For $\epsilon$ the uncertainty is $0.36$ and for $H$, $0.31$.

\section{Morphology scaling relations}
\subsection{Model}
\label{sect:scaling_reln}
We studied how the morphological parameters evolve with redshift and luminosity.
Our procedure followed that described by \cite{Bahar22}, where we parametrised the parameters as a scaling relation with luminosity and redshift.
Additionally, we also allowed the distribution width to vary.
We restricted ourselves to redshift and luminosity evolution as these are closely connected to observables and to keep the number of analyses manageable. 
Despite the large scatter between the scaling between cluster luminosity and mass \citep[e.g.][]{Pratt09}, our results for luminosity scaling should offer insight into the scaling with mass.
We constrained the scaling relation parameters, taking into account the uncertainty and covariance between luminosity and parameter value of each cluster, the likelihood for a cluster to be detected given those parameters (the selection function) and the luminosity distribution of clusters at each redshift.

We assumed that a property of a cluster, $Y$ (e.g. concentration), is related to another quantity $X$ (here $X$, for consistency with \citealt{Bahar22}, is always $\log L_{500}$ and is in units of log~erg~s$^{-1}$ in the rest frame 0.2-2.3 keV band) and redshift by
\begin{equation}
  Y = A_{\mu} +
  B_{\mu} \left( X - X_\mathrm{piv} \right) +
  C_{\mu} \left( \log E(z) - \log E(z_\mathrm{piv}) \right),
\end{equation}
the redshift evolution is $E(z) = H(z) / H_0$, $X_\mathrm{piv}$ is the X-ray luminosity pivot value, $44$, and $z_\mathrm{piv}$ is the pivot redshift, $0.3$ (which is close to the sample mean).
$A_\mu$, $B_\mu$ and $C_\mu$ are parameters describing the scaling relation.
We assumed that the distribution about this relation has a normal distribution with width given by a similar scaling relation
\begin{equation}
\log Y_\sigma = A_\sigma +
  B_\sigma \left( X - X_\mathrm{piv} \right) +
  C_\sigma \left( \log E(z) - \log E(z_\mathrm{piv}) \right).
\end{equation}
By using $\log Y_\sigma$ we avoided negative widths, but we also imposed a minimum value on $\log Y_\sigma$ of $-1.5$, to prevent unnaturally narrow distributions.
Therefore we allowed the width and mean of the distribution to vary with both redshift and luminosity.
We normalised the probability density functions (PDFs) to ensure that the integral was unity within the integration ranges of our analysis.
Integration was done by summing over a fixed grid, using 100 bins between $X=42$ and $46$.
In the $Y$ dimension, we used 200 bins over ranges $-2 \leq c_{80-800} \leq 0$, and $-2 \leq c_{500} \leq 0$.
We used 500 bins over ranges $-7 \leq n_{50} \leq 0$, $-2.0 \leq n_\mathrm{s} \leq 2.5$, $-4 \leq \alpha \leq 2$, and $-4 \leq \alpha_{50} \leq 2$.
For $H$, $\epsilon$, and $M_1$ to $M_4$ we used 100 bins between 0 and 1.
To study the impact of allowing a change in scatter, we also repeated the analysis fixing $B_\sigma$ and $C_\sigma$ to be 0.

The joint probability of the measured values $(\hat{X},\hat{Y})$ for the observables $X$ and $Y$ is given by
\begin{multline}
  P(\hat{X},\hat{Y},X,Y,I|\theta,z,t,N_\mathrm{H}) = P(I|X,Y,z,t,N_\mathrm{H}) \\ P(\hat{X},\hat{Y}|X,Y)
  P(Y|X,\theta,z) P(X|z),
\end{multline}
where $P(I|X,Y,z,t,N_\mathrm{H})$ is the selection function (calculated from the fitted model in Appendix \ref{appen:seln}), $P(\hat{X},\hat{Y}|X,Y)$ is the measurement uncertainty on the $X$ and $Y$, $P(Y|X,\theta,z)$ is calculated from the scaling relation and its width with its parameters $\theta$.

$P(X|z)$ accounts for the cosmological distribution of our observable $X$, and is calculated using
\begin{equation}
  P(X|z) = \int_{M} P(X|M,z) P(M|z) \, \mathrm{d}M,
\end{equation}
where $P(X|M,z)$ is calculated from the \cite{Chiu22} luminosity-mass scaling relation (their equation 67 and its measured width) and $P(M|z)$ is calculated from the \cite{Tinker08} mass function.

To calculate the likelihood for a single cluster we marginalised over the nuisance variables $(X,Y)$, to give
\begin{multline}
  P(\hat{X},\hat{Y},I|\theta,z,t,N_\mathrm{H}) = \int \int_{X,Y} P(I|X,Y,z,t,N_\mathrm{H}) \\
  P(\hat{X},\hat{Y}|X,Y) P(Y|X,\theta,z) P(X|z) \, \mathrm{d}X \mathrm{d}Y.
\end{multline}
To account for the cluster is detected given observables $X_i$ and $Y_i$ we used Bayes theorem, giving the likelihood for cluster $i$
\begin{multline}
  \mathcal{L}(\hat{X}_i,\hat{Y}_i|I,\theta,z_i,t_i,N_{\mathrm{H},i}) = \\
  \frac{ P(\hat{X},\hat{Y},I|\theta,z_i,t_i,N_{\mathrm{H},i}) }
  {\int \int_{\hat{X}_i,\hat{Y}_i} P(\hat{X},\hat{Y},I|\theta,z_i,t_i,N_{\mathrm{H},i}) \, \mathrm{d}\hat{X} \mathrm{d}\hat{Y}}.
\end{multline}
The total log likelihood is then computed by summing the log likelihood for all the clusters in the sample.

\begin{figure*}
  \centering
  \includegraphics[width=0.44\textwidth]{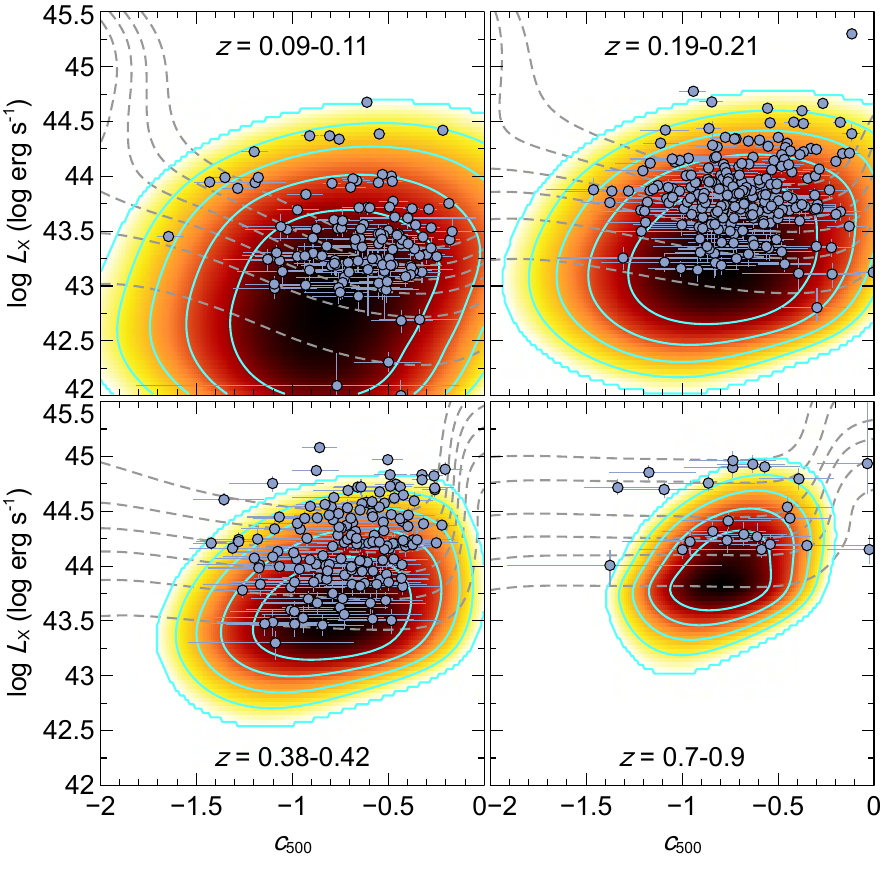} \hspace{0.5cm}
  \includegraphics[width=0.44\textwidth]{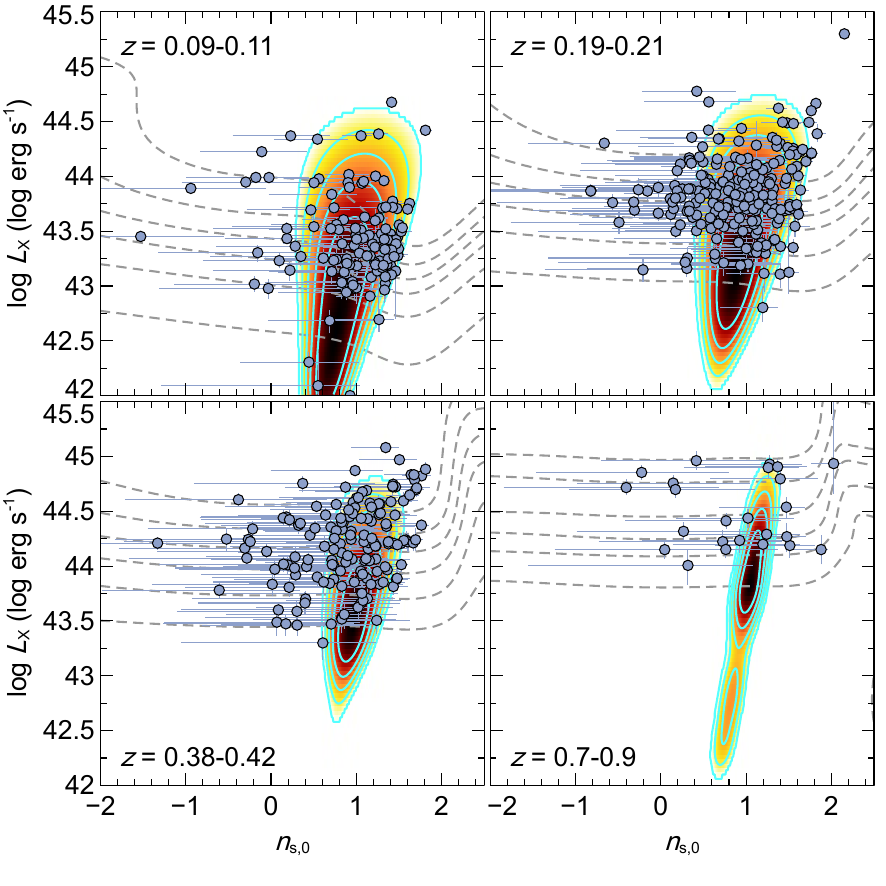}
  \caption{
    Clusters with $>50$ counts plotted on PDFs combining the maximum likelihood scaling relation, selection function and mass function.
    Shown are the clusters in four redshift bins, plotting the concentration, $c_{500}$ (left), and scaled density, $n_{\mathrm{s},0}$ (right).
    The images show the average maximum likelihood scaling relation combined with the selection function and mass function for the clusters in the redshift ranges, with the solid contour lines at difference levels of $-2$, $-4$, $-6$, $-8$ and $-10$ from the maximum.
    The dashed contour lines show the average selection function, at levels of 0.05, 0.2, 0.4, 0.6, 0.8 and 0.95.
  }
  \label{fig:pdf_pts}
\end{figure*}

While we followed the method of \cite{Bahar22}, there were some differences in our analysis.
Our selection function also accounted for the morphological parameter being studied (except for the 2D shape parameters, like ellipticity).
We allowed the width of the scaling relation to vary both with redshift and luminosity.
The joint measurement uncertainties on the morphological parameter and luminosity were calculated from the MCMC chains, and binned using our integration grid to produce  $P(\hat{X},\hat{Y}|X,Y)$.

\begin{figure*}
  \centering
  \includegraphics[width=0.9\textwidth]{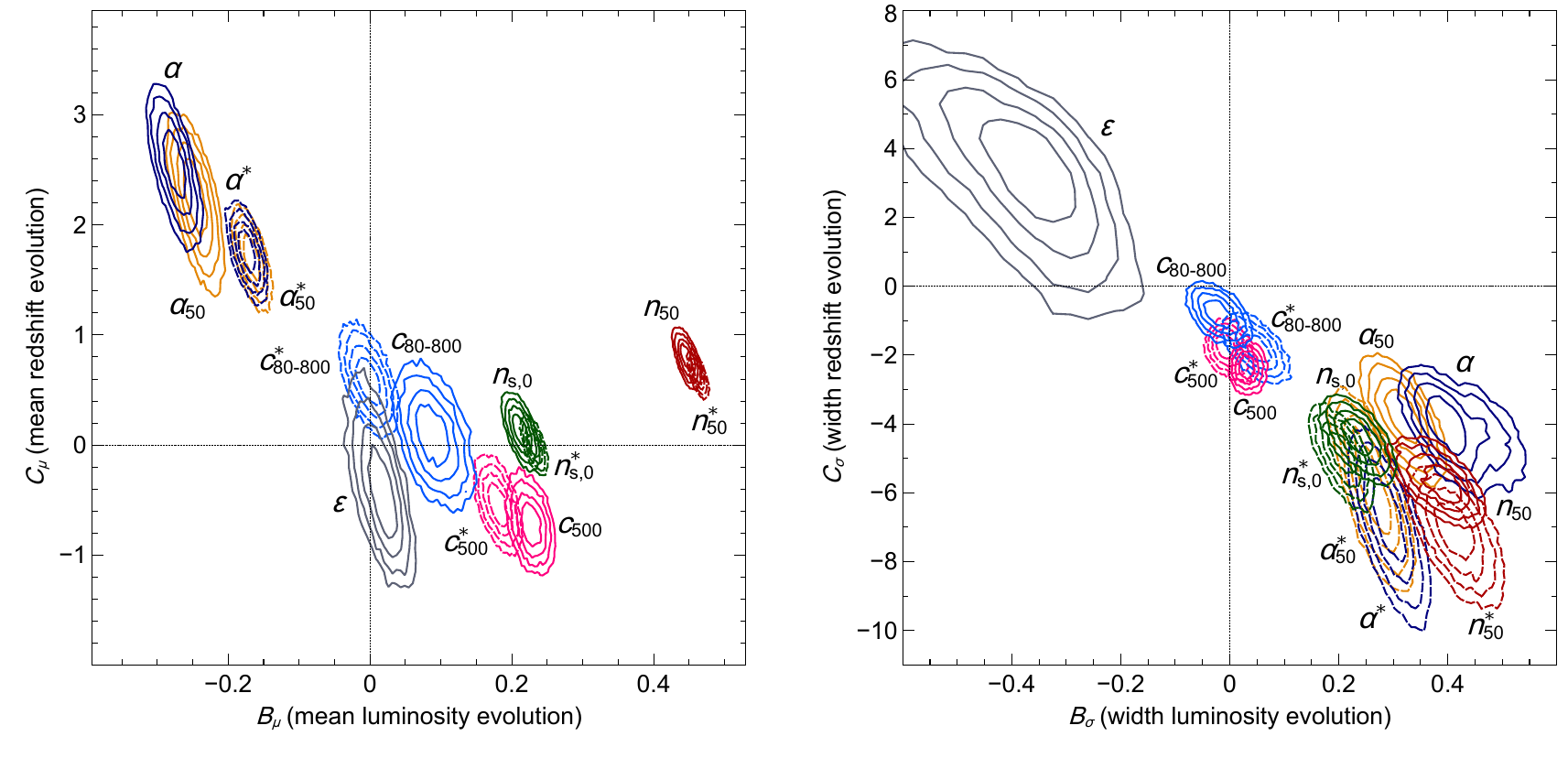}
  \caption{
    Evolution of mean and width for several parameters.
    (Left) The confidence contours for the evolution of $\mu$, with the results using the cluster fit positions as solid lines and the peak-centred cluster results as dashed lines.
    (Right) The confidence contours of the evolution of $\sigma$.
    The contours contain 39.3, 67.5, 86.4 and 95\% of the MCMC samples.
    See Appendix~\ref{appen:scaling} for corner plots of the other parameters.
  }
  \label{fig:evol_mean_width}
\end{figure*}

\subsection{Results}
\subsubsection{Fits and posterior probability distributions}
We ran the analysis for the scaling relations of several parameters with X-ray luminosity, the concentration ($c_{80-800}$ and $c_{500}$), inner density ($n_{50}$ and $n_{\mathrm{s},0}$) and inner density slope ($\alpha_{50}$ and $\alpha$), for both the fit and peak centred versions.
We also did the same analysis for the ellipticity ($\epsilon$) and slosh ($H$).
To demonstrate the fit to the data, we show in Fig.~\ref{fig:pdf_pts} the average maximum likelihood scaling relation combined with the selection function and mass function for the clusters, $\ln \left< P(I|X,Y,z,t,N_\mathrm{H}) P(Y|X,\theta,z) P(X|z) \right>$, and the average selection function, $\left< P(I|X,Y,z,t,N_\mathrm{H}) \right>$, compared to data points for the concentration, $c_{500}$, and scaled central density, $n_\mathrm{s,0}$.
The best fitting model appears to be a reasonable fit to the data points.
The analysis was also repeated assuming no evolution in width ($B_\sigma=C_\sigma=0$) and for the 100 count minimum subsample.

Fig. \ref{fig:evol_mean_width} shows the posterior probability contours for the evolution of the mean and its width, for both the standard and peak-centred versions of the parameters.
Tables of the values and corner plots of the other posterior parameter distributions are found in Appendix~\ref{appen:scaling}.
To better understand the likely systematic uncertainties, it is useful to compare the results for the two different cluster centres, for the analysis with fixed distribution widths, and for the smaller 100 count subsample.

\subsubsection{Concentration}
The concentration at scaled radius, $c_{500}$ has a value at the pivot point of $-0.68$, while it is $-0.62$ for $c_{500}^*$.
The model gives positive scaling with X-ray luminosity, where $B_\mu \sim 0.2$, while there is significant scaling with redshift, with $C_\mu \sim -0.6$.
The width of the distribution prefers negative redshift evolution, if our selection function and measurements are accurate, although its luminosity evolution appears statistically insignificant.
There is reasonably good agreement between the $c_{500}$ and $c_{500}^*$ results for the models with and without width evolution, and also the sample with 100 count objects, with $c_{500}^*$ having a larger $A_\mu$ and smaller $A_\sigma$ than $c_{500}$.

For fixed radius concentration, $c_{80-800}^*$ prefers a pivot value of the mean around 0.1 dex larger than $c_{80-800}$.
The analysis does not find convincing evidence for its evolution.
$c_{80-800}^*$ is consistent with no luminosity evolution of the mean, while $c_{80-800}$ is consistent with no redshift evolution of it.
Any luminosity evolution is smaller than that of $c_{500}$.
The evolution of the width appears mildly significant with negative redshift.

\subsubsection{Central density}
The central density at scaled radius, $n_{\mathrm{s},0}$, shows rather good consistency between its evolutionary parameters for the different analyses, with its $B_\mu \sim 0.2$ being very similar to that of $c_{500}$.
There is no evidence for mean redshift evolution.
The width, however, prefers a positive evolution with luminosity and a negative evolution with redshift, both for  $n_{\mathrm{s},0}$ and  $n_{\mathrm{s},0}^*$.
In contrast, for $n_{50}$ the model prefers a luminosity evolution which is twice that $n_{\mathrm{s},0}$ ($B_\mu \sim 0.45$), with also positive redshift evolution.
The width of the distribution, if allowed to evolve, again prefers to increase with luminosity and decrease with redshift.

\subsubsection{Cuspiness}
The two cuspiness parameters, $\alpha$ and $\alpha_{50}$ show very similar results.
These parameters evolve negatively with luminosity, in contrast with the other parameters.
They also show rather strong positive redshift evolution.
Like $n_{50}$ the width of the distribution evolves positively with luminosity and negatively with redshift.
We note, however, that the central slope is a more difficult quantity to measure than the density or concentration and is more likely affected by the modelling procedure.

\subsubsection{Ellipticity and slosh}
The ellipticity scaling relation shows a mean of $0.77$, with no evidence for evolution with luminosity or redshift.
This mean value is very close to the median ellipticity found in a set of simulations and XMM-Newton observations \citep{Campitiello22}.
The width, however, shows negative evolution with luminosity and positive evolution with redshift, although its uncertainties are large.
The slosh parameter prefers a slightly negative or zero mean, although this is affected by the prior of the parameter.
Its evolution is consistent with zero in both luminosity and redshift.

\section{More complex distributions}
\label{sect:complex_dist}
The above scaling relation analysis assumed that the distribution of a parameter is normal at each redshift and luminosity.
We could have introduced more complex distributions in an evolutionary analysis, but their interpretation becomes complex.
Therefore to investigate more complex distributions, we examined clusters in bins of luminosity and redshift and assumed that the parameter distributions are constant within these bins.

\begin{table*}
  \caption{Priors used for fitting distributions in redshift and luminosity bins.}
  \centering
  \begin{tabular}{lll}
    \hline
    Model & Morphological parameter & Priors \\ \hline
    Normal & $c_{500}$, $c_{80-800}$ & $\mu \sim \mathcal{U}(-2.2,0.2)$, $\log \sigma \sim \mathcal{U}(-1.3,2.0)$ \\
           & $n_{\mathrm{s},0}$ & $\mu \sim \mathcal{U}(-0.8,2.8)$, $\log \sigma \sim \mathcal{U}(-1.3,2.0)$ \\
           & $n_{50}$ & $\mu \sim \mathcal{U}(-7.7,0.7)$, $\log \sigma \sim \mathcal{U}(-1.3,2.0)$ \\
           & $\alpha$, $\alpha_{50}$ & $\mu \sim \mathcal{U}(-4.6,2.6)$, $\log \sigma \sim \mathcal{U}(-1.3,2.0)$ \\
           & $\epsilon$, $H$, $M_1$-$M_4$ & $\mu \sim \mathcal{U}(-0.1,1.1)$, $\log \sigma \sim \mathcal{U}(-1.3,2.0)$ \\ \hline
    Skew  & All & Same as Normal model, but with additional parameter skew $K \sim \mathcal{N}(0,10)$ \\ \hline
    $2\times$ Normal  & All & Two components with normal model priors, plus additional fraction $f \sim \mathcal{U}(0,1)$ \\ \hline
    Interpolated & $c_{500}$, $c_{80-800}$ & Eight points between $-2$ and $0$, with normalisation $N_i \sim \mathcal{N}(0, 4)$ on each \\
                 & $n_{\mathrm{s},0}$ & Eight points between $0$ and $2$, with normalisation $N_i \sim \mathcal{N}(0, 4)$ on each \\
                 & $n_{50}$ & Eight points between $-4$ and $-1.5$, with normalisation $N_i \sim \mathcal{N}(0, 4)$ on each \\
                 & $\alpha$, $\alpha_{50}$ & Eight points between $-1$ and $2$, with normalisation $N_i \sim \mathcal{N}(0, 4)$ on each \\
                 & $\epsilon$, $H$, $M_1$-$M_4$ & Six points between $0$ and $1$, with normalisation $N_i \sim \mathcal{N}(0, 4)$ on each \\
    \hline
  \end{tabular}
  \tablefoot{
    $\mathcal{N}(\mu,\sigma)$ is a normal distribution with mean $\mu$ and width $\sigma$.
    $\mathcal{U}(a, b)$ is a uniform distribution between $a$ and $b$.
    The interpolated model points are separated uniformly in the range specified, with the endpoints at the extreme values.
  }
  \label{tab:distprior}
\end{table*}

In this analysis we used four different distributions, a normal distribution, a skew normal distribution, a dual normal, and an interpolated model.
Priors of the parameters are listed in Table \ref{tab:distprior}.
The normal distribution model has the mean ($\mu$) and width ($\sigma$) as free parameters.
The skew normal distribution has an additional parameter for the skew $K$.
The dual normal distribution consists of two normal components.
In addition to their means and widths, the components have relative weights $f$ and $1-f$, where $f$ is a free parameter between 0 and 1.

For the interpolated model, we chose a set of six or eight parameter values which are equally separated in parameter space.
The log distribution normalisations at these values were free parameters.
Points at intermediate values were calculated using cubic spline interpolation in log space.
The overall distribution was normalised to create a PDF.
As some parameters are only allowed to be in a given range, we renormalised the PDFs for these parameters and made them 0 outside this range (e.g. 0 to 1 for $\epsilon$ and $H$ and $-2$ to $0$ for $c_{500}$ and $c_{80-800}$, with the same integration grid as in Section \ref{sect:scaling_reln}).

The objects studied were in five bins of luminosity and five bins of redshift, fitting the objects in the cosmology sample with more than 50 counts (Table \ref{tab:subsamples}).
We derive posterior probability distributions and the Bayesian evidence using the \texttt{UltraNest}\footnote{\url{https://johannesbuchner.github.io/UltraNest/}} package \citep{Buchner21}.
The detailed resulting parameter distributions can be found in Appendix~\ref{appen:distn}.

\begin{figure*}
  \centering
  \includegraphics[width=\textwidth]{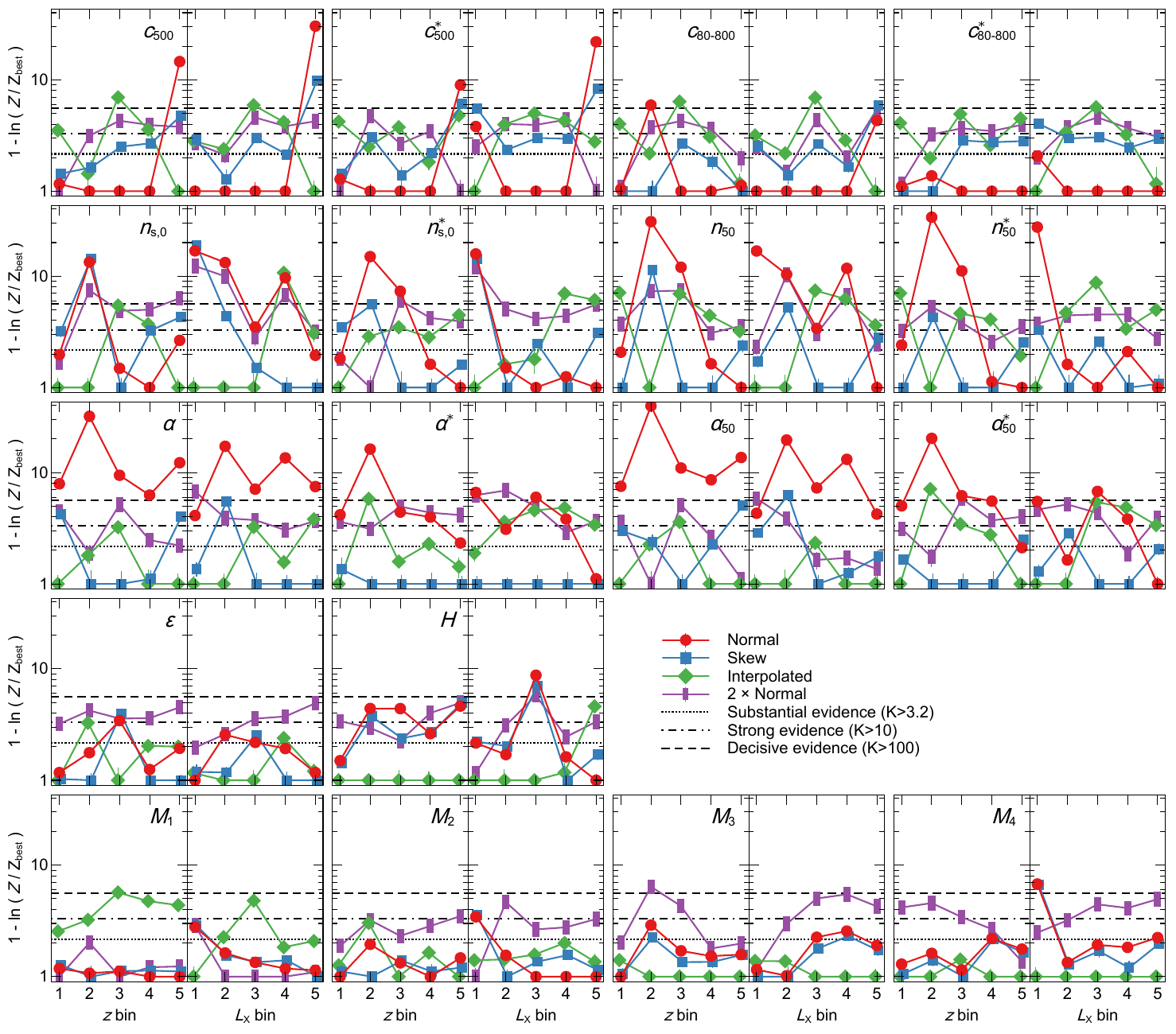}
  \caption{
    Comparison of model distributions in bins of redshift and luminosity.
    Shown is the difference in Bayesian evidence (the Bayes factor) between each model and the model with the largest Bayesian evidence, where the models with the largest evidence are plotted at lower values and the best model has a value of $1$.
    Shown are typical thresholds for Bayes factors from \cite{KassRaftery95}.
  }
  \label{fig:evidence}
\end{figure*}

\begin{figure*}
\includegraphics[width=\textwidth]{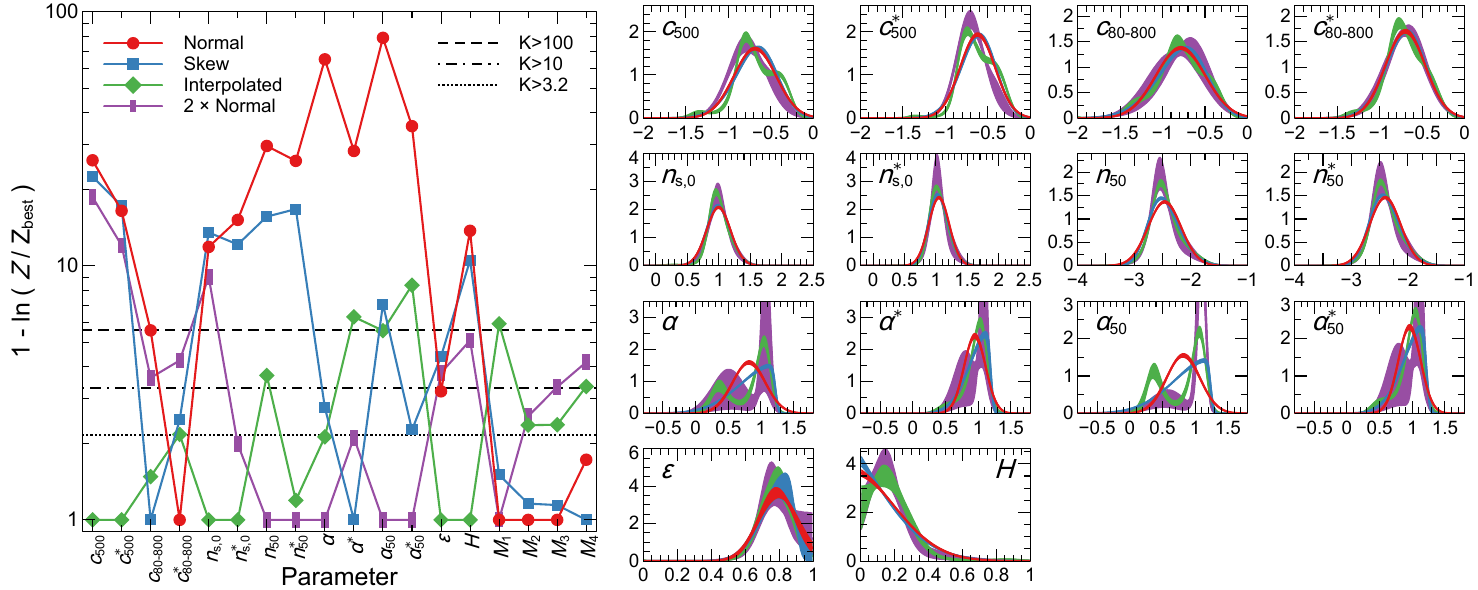}
\caption{
Comparison of different distribution models for the whole sample.
(Left panel) Model comparison, as in Fig.~\ref{fig:evidence}, for each parameter, assuming that the clusters have a single distribution.
Models with larger evidence have lower values.
(Right panels) The PDFs of each parameter, for each of the model types, coloured as in the left panel. The shaded regions contain 68.27\% of samples.
}
\label{fig:dist_all}
\end{figure*}

The standard method for doing model comparison in the Bayesian framework is to compute a Bayes factor, $K=P(D|M_\mathrm{a})/P(D|M_\mathrm{b})=Z_\mathrm{a}/Z_\mathrm{b}$, given data $D$ and the two models to compare $M_\mathrm{a}$ and $M_\mathrm{b}$.
The marginal likelihood for the model, the `evidence' $Z$, is difficult to compute using MCMC sampling or directly, but can be calculated using nested sampling.
In Fig.~\ref{fig:evidence} we show the Bayes factor between the models in the bins of redshift and luminosity for the concentration, central density, cuspiness, ellipticity, slosh and multipole magnitudes, for both the average centred and peak centred results.
The comparison is between each model and the best model with the largest evidence for that particular bin.
In order to be able to plot using a log scale, we invert the scale and plot $1-\ln (Z/{Z}_\mathrm{best})$, so that the best model is plotted as 1 and those with less evidence have higher values.
Included in the plot are typically used thresholds for changes in evidence, which can be used for ruling out models.
For example if a point lies above the line for `Strong evidence', then it has little evidence supporting it, while points which lie below the line for `Substantial evidence' have similar levels of evidence to the best model.

Model comparison using Bayes factors favour simpler models with fewer parameters.
However, the chosen priors can strongly affect the results.
Here we use non-informative priors to use the data to constrain the average value of the parameter, and given that we do not have any advance knowledge of their distribution shape.
There is risk that diluting the parameter space with a too broad non-informative prior may over penalise models, and so we emphasise that our comparisons are between the models combined with their priors.

Examining the concentration distributions, we find that a normal distribution is statistically acceptable in most of the redshift and luminosity bins.
Exceptions include the highest luminosity and redshift bins for $c_{500}$ or $c^{*}_{500}$, which prefer dual normal or interpolated models, and for a couple of bins for $c_{80-800}$, although these are not seen for $c_{80-800}^*$.

For the scaled density, $n_{\mathrm{s},0}$, a normal distribution is often not acceptable, except for high luminosity and redshift $n_{\mathrm{s},0}^*$ bins.
Instead, a mixture of interpolated, skew and dual models are preferred.
A very similar picture is also seen for $n_{50}$, with the similar pattern in the same bins.
Examining the model distributions (Appendix~\ref{appen:distn}), the models appear asymmetric with a sharper edge to one side of the distribution and a narrower core, than a normal distribution would prefer.
Typically tails are seen to higher densities in the redshift bins and lower densities in the luminosity bins.
For all these density parameters, the peak position evolves with luminosity, similarly to the scaling relation results, while for $n_{50}$ evolution with redshift is also visible.

The cuspiness parameters $\alpha$ and $\alpha_{50}$ show similar results to each other.
Almost none of the bins find the normal model acceptable.
The distributions (Appendix~\ref{appen:distn}) with a fitted cluster centre ($\alpha$ and $\alpha_{50}$) show two peaks, while those using the cluster peak ($\alpha^*$ and $\alpha^*_{50}$) instead show a skewed distribution, although the differences are not obvious in Fig.~\ref{fig:evidence}.

The ellipticity does not generally prefer any particular model for the distribution, although the normal distribution is roughly acceptable for most of the bins, although the skew or interpolated model is often better.
For slosh, an interpolated model is preferred for in most bins.
The picture for $M_1$ to $M_4$ is more complex, although a normal distribution is acceptable for many bins.
For $M_3$ and $M_4$ an interpolated model is favoured.

One can also use the assumption that the distributions are the same for the whole sample, with no evolution with redshift or luminosity.
Figure~\ref{fig:dist_all} compares the Bayes factors of the resulting models for each parameters and shows their PDFs.
Concentrations $c_{500}$ and $c_{500}^*$ both prefer complex interpolated models, with two peaks and heavily disfavour a normal distribution, while the evidence for complex distributions is weaker for $c_{80-800}$ and $c_{80-800}^*$.
The density parameters prefer interpolated or dual-normal models, and disfavour normal models.
The central slope parameters also are inconsistent with normal models, preferring dual normal or skewed models, with less evidence for a secondary peak if using the cluster peak centre.
Both ellipticity and slosh prefer an interpolated model over a normal model, while the multipole magnitudes are consistent with normal distributions.

\section{Discussion}
\label{sect:discuss}
\subsection{Comparison with other samples}
\label{sect:other_samp}

\begin{figure*}
  \includegraphics[width=\textwidth]{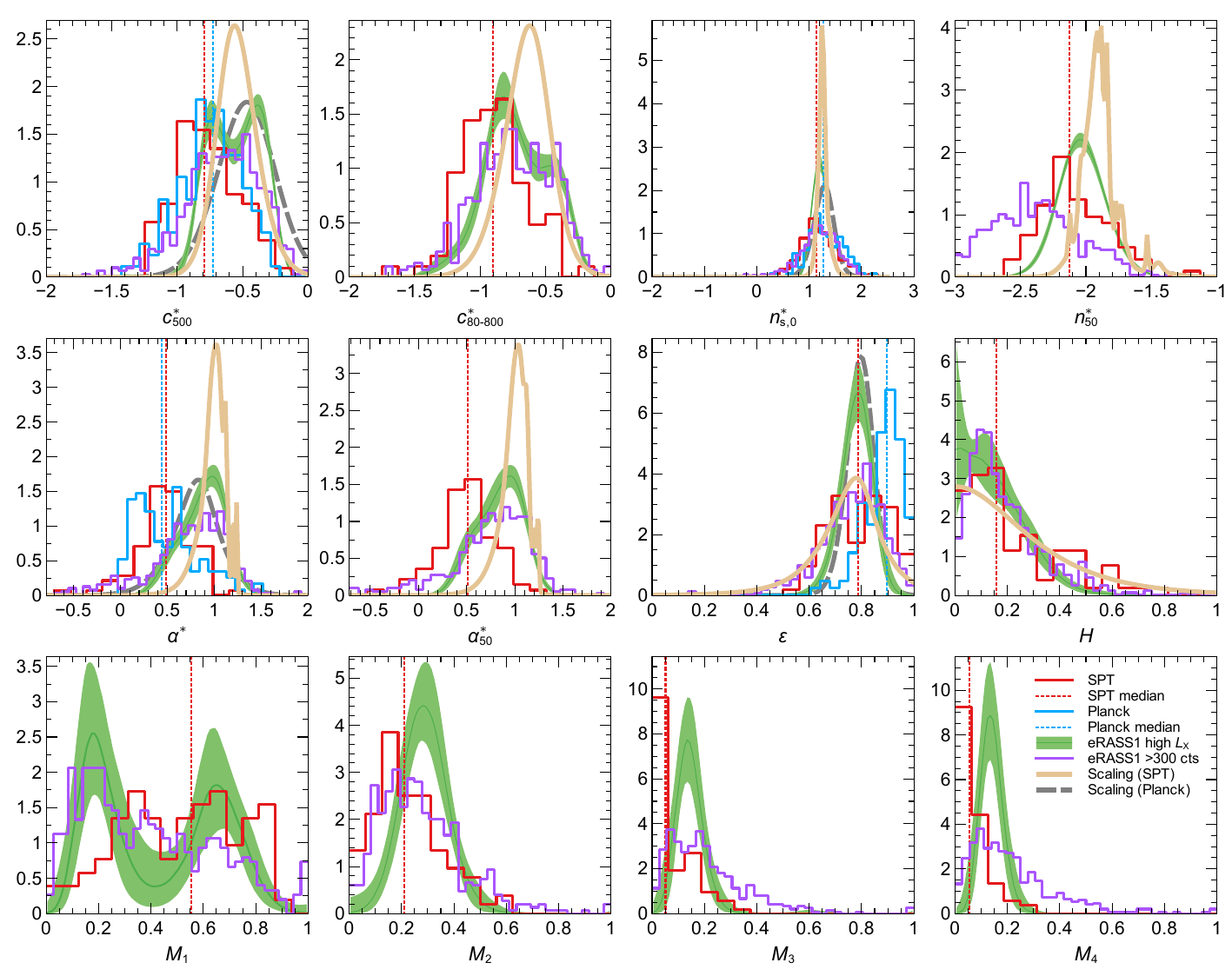}
  \caption{
    Comparison of eRASS1 probability densities compared to SPT and Planck selected samples.
    The distribution of the parameters for the Chandra-SPT sample is plotted as a histogram, with the median value indicated by a dashed line.
    The Planck values are for ESZ clusters from \cite{Lovisari17}.
    The model distribution for the eRASS1 clusters in the high luminosity bin is plotted ($\log L_X=44.3-45.6$, fitted using interpolated model).
    For those parameters where we fitted a scaling relation (Appendix \ref{appen:scaling}), we computed matched PDFs given the redshift and luminosities of the SPT and Planck samples.
    Also plotted are the distributions of parameter values for the bright 300 count cluster eRASS1 subset.
  }
  \label{fig:other_compare}
\end{figure*}

In Fig.~\ref{fig:other_compare} we compare the results for our analysis of eRASS1 clusters with our analysis of 83 clusters from the SZ (Sunyaev-Zel'dovich) effect selected South Pole telescope (SPT) sample observed by Chandra.
In addition, we show the distributions for 120 clusters from the Planck early SZ (ESZ) sample, measured using XMM-Newton data by \cite{Lovisari17}.
The cluster models used to fit the Chandra data are the same as for our eRASS1 analysis, although the typical exposure times are much longer for the Chandra data and the spatial resolution of the telescope is much better.
For this analysis of the radial properties, the centre of the cluster was fixed at the peak of the emission, as in \cite{Sanders18}.
The Chandra analysis is discussed further in Paper I.
We note that for this analysis the masses and radii for the SPT clusters were taken from the SPT-data \citep{Bleem15}, rather than derived from the Chandra data.
For the XMM-Newton observed clusters we used Planck radii and masses to convert their physical density at $0.02R_{500}$ to $n_{\mathrm{s},0}$, as in Paper I.

For the eROSITA data we plot the interpolated model distributions in our highest luminosity bin from the analysis in Section \ref{sect:complex_dist}, which is roughly the luminosity range of the SPT sample.
As the Chandra and XMM-Newton measurements are fixed to the cluster peak, we show the peak based measurements for eROSITA.
We also include a direct histogram of the values for the subsample of eRASS1 clusters with more than 300 counts.
For those parameters where we fitted scaling relations, we compute eRASS1 effective matched PDFs for the SPT and Planck samples, given our scaling relations and the individual X-ray luminosities and redshifts of the SPT and Planck ESZ clusters, where ESZ cluster luminosities were scaled from the values in \cite{Lovisari20}.
Planck clusters typically lie at lower redshift than SPT clusters and have a different luminosity/mass distribution, leading to differences between the PDFs.

In Paper I we compare the distribution of parameters in different eRASS1 subsamples with SPT and Planck, finding that the eRASS1 clusters are more concentrated.
The result here is similar to that found for the raw concentration distributions, even though the effect of selection is accounted for.
For $c_{500}$, the difference between the eRASS1 SPT-matched PDF and the SPT median
is around $0.23$, or 70\% more flux on average in the inner aperture for eRASS1 objects compared to SPT.
The difference between eRASS1 Planck-matched PDF and Planck median is around $0.25$, or 78\% more flux inside $0.1 R_{500}$.
We note, however, that there are differences between the concentrations measured for the same clusters.
In Paper I, we find that the median measurements for $c_{500}$ are 0.1 dex larger for eRASS1 compared to values from Chandra for SPT clusters and XMM-Newton for Planck clusters.
Reasons for the differences can include different point source masking and unresolved point sources, cluster radii, cluster backgrounds, PSF calibration and the larger eROSITA field of view.
This accounts for around half the difference, giving concentrations larger by $0.13$-$0.15$ in eRASS1 compared to the SZ-selected objects after this is subtracted (around 35 to 41\% more flux).

Despite the differences seen in the concentration, we see very little difference between the peak central density $n_{\mathrm{s},0}^*$ for eRASS1 and the SZ samples.
The peak scaling relation density lies almost exactly at the same value as the Planck cluster value.
eRASS1 clusters are found to have around 15\% lower fluxes than Chandra and XMM (B24; \citealt{Migkas24}), although the effect of this on the density should be small ($\sim 0.03$ dex).
The difference between SPT clusters and eRASS1 clusters is larger for the density at fixed radius, showing a difference of 0.2 dex higher densities at 50~kpc.
Although the central density showing reasonable agreement at scaled radius, if the density slope or cuspiness is measured compared to SZ-cluster measurements, we find values around 0.5 larger for eRASS1 clusters.

eRASS1 clusters seem to be more concentrated than for SZ samples, despite modelling selection effects from the underlying population.
Similarly, the clusters are more steeply peaked in the X-ray band, but in contrast the scaled central densities are consistent.
Explanations for the difference in concentration may be that we are either looking at intrinsically different cluster populations, our selection function modelling is inaccurate, or there are differences in observational effects for the samples.

We measured the properties for the sample as a whole assuming a selection function.
The SPT and Planck clusters were not selected in the X-ray waveband, and so should not have a bias towards a cool core.
However, we note that both these SPT and Planck selected samples are not pure SZ-selected samples, as not all objects above some detection threshold were X-ray observed, which could potentially bias their results.
In addition, for Planck, we are using the subsample where $R_{500}$ fits within the XMM field of view, containing 120 out of 189 SZ-selected objects.
SZ samples may also be affected by different selection on properties connected to morphology.
For example, if they preferentially detect unrelaxed or irregular systems, this could lead to them being less concentrated.
\cite{AndradeSantos17} find a lower fraction of cool core objects in a Planck SZ selected sample compared to an X-ray flux selected sample.
\cite{Rossetti17} find a similar result, comparing Planck SZ selected and X-ray selected samples.
\cite{Lovisari17} confirm that a Planck-selected sample tend to be more morphologically disturbed compared to X-ray selected surveys.
It is claimed that the matched filter technique used to detect clusters in SZ surveys is independent of cluster astrophysics \citep[e.g.][]{Melin05}.
However, simulations of Planck data find clusters with steeper pressure profiles produce more complete samples compared to a standard gNFW set of profiles \citep{Gallo24}.
There are differences in the Planck selection found for different ellipticities, although this effect is relatively small except for objects with large angular size.
Cluster mergers may produce shocks, which result in pressure jumps, to which SZ selection might be more sensitive than X-ray selection \citep[e.g.][]{Ruan13}, similar to the effect of cool cores in X-ray selection.
Simulations predict that the scatter in the SZ $Y-M$ relation is caused by dynamical state \citep[e.g.][]{Battaglia12}.

It may be that our selection function is not valid and we are not correcting for the observational effects properly.
If profiles of the models generated for the selection are very different to real clusters, or their 2D shape is sufficiently different, this could change their detection efficiency.
We also repeated the analysis of the distribution in bins of redshift and concentration, following Section \ref{sect:complex_dist}, but assuming that the selection function is unity.
In the case of the Gaussian model, the average Gaussian distribution mean in each bin of redshift or luminosity changes by 0.05 or less, for both $c_{500}^{*}$ and $c_{80-800}^{*}$.
We also separately fitted the evolution of $>50$ and $>100$ count subsamples in Section \ref{sect:scaling_reln}, finding results which were reasonably similar, despite having different selection functions.
Therefore, we would have to make large changes to the selection function to change the results significantly.

The density profiles measurements could also be biased in some way, for example, if clusters have profiles sufficiently different from the fitted functional form, or if the priors are affecting the result.
However, the $>300$ count subsample shows a similar result, which should be much less affected by priors.
For the faintest clusters, the choice of the X-ray peak as the cluster centre could also bias the concentrations upwards due to Poisson fluctuations.

If there is a substantial fraction of contaminating non-clusters, such as active galactic nuclei (AGN), these will likely have high concentrations and bias the distribution.
Contaminating AGN could also affect a subset of the objects by artificially increasing their concentration values, although the cosmology sample has a purity of 95\% (B24).
We restrict our distribution analysis to the cosmology sample, further selecting  clusters with more than 50 counts, which increases the median extension likelihood from $12.0$ to $19.3$ and likely increasing the purity beyond 95\%.
The substantial fraction of high concentration objects cannot be purely due to contamination if $<5$\% of clusters are not real.
We also still see similarly high concentration values for our $>300$ count subsample which should be purer than than the full sample and less affected by unresolved AGN.

However, we see differences between the same set of clusters observed by eROSITA and XMM-Newton or Chandra, where eRASS1 $c_{500}$ concentrations are around 0.1 dex larger (Paper I).
If the cause of this difference is more important for the population of clusters less represented in SPT or Planck surveys, this could give rise to the 0.1-0.2 dex further differences between the whole populations.
For example, if high concentration objects had their concentrations strongly boosted by AGN contamination, this could give rise to such a difference.

As noted in Paper I, the 2D shapes of clusters are rather similar in the eRASS1 and SPT samples.
The median SPT ellipticity value ($0.79$) lies extremely close to the one from eRASS1, with the eFEDS distribution being similar.
The distribution of slosh and $M_1$ to $M_4$ are are also rather similar between the samples.
There we find that X-ray selected clusters have a rather similar 2D shape to SPT-selected clusters.
This is somewhat surprising, as it might be expected that X-ray surveys detect more cool-core, and therefore more regular clusters in terms of 2D shape, compared to SZ surveys \citep[e.g.][]{Maughan12}.
We note, in contrast, that \cite{Nurgaliev17} finds no difference between X-ray and SZ-selected samples, which may be because its 400d X-ray sample \citep{Burenin07} does not have a pure X-ray flux selection.
In addition, the Planck clusters have high values of $\epsilon$, meaning that they are much rounder than eRASS1 and SPT.
This is likely due to measurement differences, as the matched sample of eRASS1 and Planck clusters (Paper I) show the XMM results have significantly higher values of $\epsilon$ than our eRASS1 analysis.
This could be because the XMM results are sensitive to a different part of the cluster, or the measurement is affected by the much larger number of counts.
The eRASS1 multipole magnitude high count distributions and fitted models look rather similar to the distributions for the SPT clusters, although the SPT fits prefer values closer to zero for $M_3$ and $M_4$.
These multipole magnitude comparison is less clear as the statistical measurement uncertainties are large and the parameters cover a limited range of values.

\subsection{Cluster position}
\label{sect:posn}
As noted previously \cite[e.g.][]{Sanders18}, the choice of centre of a cluster has a large impact on some parameters such as the central density.
We tried two methods in this analysis - fitting the cluster position in the analysis and forcing a peak position.
However, we note that our position of the peak is determined from a smoothed map, which can be dependent on the data quality and smoothing scale (which is a fixed angular scale and is therefore redshift dependent).
We find that using the peak positions increases the average concentrations by $0.063$ and $0.010$ for $c_{500}$ and $c_{80-800}$, respectively, from the evolutionary analysis.
In Paper I, figure 18, we show that there is increased bias in the measured parameters for objects with low number of counts when using peak compared to average positions.
For example, for a 50 count cluster, our minimum here, the peak measured concentrations can be biased upwards by $0.04$-$0.05$ dex, while the position fitted concentrations have little bias.
Therefore, $2/3$ to $1/2$ of the measured difference could be due to this bias, as many of the clusters are close to the 50 count limit, although there may be a real differences caused by non symmetric clusters.
Similarly, using the peak position increases the inner slopes by $0.12$.
The central density $n_{50}$ increases by $0.037$ and $n_{\mathrm{s},0}$ by $0.046$, much of which is likely due to the centre bias.
Future analyses using deeper eROSITA surveys will be able to test in more detail the differences between choice of centre as we will have larger samples with more counts.

\subsection{Evolution of the inner gas properties}
Our modelling of the scaling relations allows us to determine the evolution of several of our cluster parameters, under the assumption that they have normal distributions, their evolution can be modelled by a scaling relation, and our selection functions properly model the selection of real clusters, and that the assumed cosmology is correct.

One of the most interesting cluster morphological properties is concentration, as it is easy to measure and sensitive to cool cores.
On average 16-20\% of the cluster luminosity is emitted from the inner 80~kpc (depending on how the cluster centre is defined), and 21-24\% is emitted from the inner $0.1 R_{500}$.
The evolution of the concentration using a fixed aperture ($c_{80-800}$ and $c_{80-800}^*$) shows differences depending on how the cluster centre is chosen, suggesting that we do not measure significant evolution due to systematic uncertainties. 
Alternatively, the peak position might be intrinsically better for measuring the properties of a cool core, although one must be aware of bias (Section \ref{sect:posn}).
A lack of evolution is in contrast to the picture that cool core clusters have generally higher luminosities \citep[e.g.][]{EdgeStewart91,Pratt09,Hudson10,Mittal11b}.
However, the concentration in a scaled aperture, $c_{500}$ and $c_{500}^*$, both show positive luminosity evolution, in agreement with the picture of cool cores having higher luminosities.
We also find some evidence for lower redshift clusters having larger $c_{500}$.
The evolution in the widths of the concentration distribution is relatively small.

The weaker luminosity evolution of $c_{80-800}$ compared to $c_{500}$ may be due to cool cores evolving more similarly with clusters if measured in a fixed physical aperture.
\cite{McDonald17} previously examined a sample of massive cool core objects, finding that the average over-density profile did not evolve in redshift.
However, we are not looking at cool cores in particular, but the whole cluster population.
When looking at the evolution of cool cores, the peak-centred quantities $c_{500}^*$ and $c_{80-800}^*$ are likely more robust, as the other parameters could be affected by the shape of the surrounding ICM.
However, peak-centred parameters have additional bias in fainter clusters (Section \ref{sect:posn}).

Examining the central density, we find that the physical gas density at a physical radius of 50~kpc ($n_{50}$ and $n_{50}^*$) evolves strongly positively with luminosity.
The scaled density ($n_{\mathrm{s},0}$ and $n_{\mathrm{s},0}^*$), which accounts for self-similar evolution, shows luminosity evolution similar to $c_{500}$ and no redshift evolution.
These are consistent with the picture that cool core clusters have denser cores and are more luminous.
The width of the distributions increases with luminosity and decreases with redshift.

The central density slope or cuspiness, $\alpha$ and $\alpha_{50}$, show negative luminosity evolution, but positive redshift evolution.
The widths of their distributions decrease with increasing redshift and increase with luminosity.
The slope parameters do not appear to evolve in a similar way to the central density, suggesting a change in the shape of profiles.
The negative evolution with luminosity may suggest that more massive clusters have a more active merger history.
Another possibility is that the non-normal distribution (Section \ref{sect:complex_dist}), distorts the evolutionary parameters.

\subsection{Evolution of the 2D shape}
Our simulations suggest that the 2D cluster shape (e.g. ellipticity or slosh) does not significantly affect the detection of clusters for reasonable distortions (Paper I).
In bins of redshift and luminosity ellipticities are clustered around values of 0.8  (Appendix \ref{appen:distn}).
The slosh parameter, $H$, distribution peaks at values of $0.1-0.2$.
The evolution analysis shows the mean ellipticities is around 0.78, with no significant evolution with redshift or luminosity.
However, the width of the distribution shows negative evolution with luminosity (i.e. more luminous clusters have a narrower ellipticity range), but the width of the distribution increases with redshift.
Therefore there appear to be more extreme objects at lower luminosities (i.e. lower masses) and higher redshifts.

Similarly, our slosh parameter, designed to measure asymmetries similar to a sloshing cold front, has a peak towards zero slosh in the evolution analysis.
It also shows little evidence for a shift in peak with redshift and luminosity.
Similar to ellipticity, the width of the distribution does reduce with increasing luminosity and increases with redshift.

The lack of evolution in these parameters is suggestive that the average cluster over redshift and luminosity has a similar rate of high morphological disturbance.
The width of the distribution reduces with increased redshift and luminosity, however.
Although we did not find a strong effect of 2D shape on selection itself, if it is correlated with other parameters such as concentration, we may be missing very concentrated but spherical objects, or non-concentrated highly disturbed objects.
This effect of correlation between parameters and the selection should be less important for the profile-based parameters, as the profiles were generated based on a model obtained from real cluster profiles.

\subsection{Single or multiple populations}
One often studied topic is whether the cluster population consists of a single continuous population of objects or whether there are different subpopulations of objects leading to bimodal distributions \cite[e.g.][]{Pratt10,Santos10,Hudson10,Sanders18,Ghirardini22,Riva24}.
Whether one sees a single population of objects or sees bimodality can also depend on the parameter space being examined.
There is a large multidimensional space of possible cluster properties, where clusters can look alike if viewed along certain projections, but they might not if looked at in other ways.
For example, the presence of a cool core may not be a useful indicator of whether there is a merger taking place \citep[e.g.][]{Hudson10,Lovisari17}.
Bimodality is usually assessed by examining the distribution of a particular quantity to see whether more than one normal component is necessary.

It has been known for a long time that a large fraction of X-ray selected galaxy clusters have mean radiative cooling times which are relatively short in their centres \citep[e.g.][]{Bauer05}, which are known as cool core clusters.
These objects are seen as having steeply peaked surface brightness profiles compared to the flatter profiles in other systems.
Cool core clusters also show evidence for multiphase material in their centres and evidence for AGN feedback \citep[e.g.][]{Fabian12}.

However, it is not clear whether cool core to non cool core clusters is a simple continuum, or whether there are distinct populations.
Strong cool core objects are characterised by a very concentrated X-ray surface brightness profile, low central mean radiative cooling time (or low entropy), reductions in temperature in the very central region and often the X-ray peak and brightest cluster galaxy being at the same location.
Several of the morphological parameters we have measured are directly sensitive to the presence of a cool core (e.g. concentration, inner density slope and central density).
\cite{Hudson10} examined the detailed profiles of their sample of nearby bright galaxy clusters and found, examining histograms of properties, that they could be split into three subsets: strong cool core, weak cool core and non cool core.
\cite{Pratt10} found that the central entropy in the REXCESS sample showed either a bimodal or skewed distribution.
\cite{Sanders18} examined the distribution of entropy in the SPT sample of galaxy clusters observed by Chandra, finding evidence for a bimodal distribution of the central entropy.

Other authors, however, have not found evidence for bimodality (or trimodality) in their clusters.
\cite{Santos10} examined the concentration distribution of different samples, finding no evidence for bimodality.
\cite{Ghirardini22} studied the distributions of morphological parameters for the eFEDS sample of clusters observed by eROSITA, finding no evidence for bimodality.
\cite{Riva24} examined the central entropy distribution in the CHEX-MATE cluster sample, finding no evidence for bimodality.

We examine the distribution for some forward-modelled parameters in bins of redshift and luminosity in Section \ref{sect:complex_dist}.
We do not find find a single model distribution which best describes all the parameter distributions.
For the ellipticity, its distribution is consistent with normal in most bins, although interpolated or skew models are preferred in a couple of bins.
In contrast, interpolated models are usually preferred for $H$, with a peak preferred in the $0.1$-$0.2$ range and a tail to higher values.
For the multipole magnitude parameters, $M_1$ to $M_4$, we find that in many bins the normal model is adequate to describe the data.

For the concentration, normal distributions are statistically reasonable, except that there are particular bins of redshift of luminosity which strongly prefer other models, such as the highest redshift and luminosity bins for $c_{500}$ and $c_{500}^*$.
However, in some of the bins (Appendix \ref{appen:distn}) the interpolated model is suggestive of peaks at similar values, although the statistical evidence in each bin is not strong.

The central density parameters show no clear model which best describes the distribution in each bin, although the normal distribution is heavily disfavoured for many of them.
The distribution plots show tails in some of the bins, typically to higher densities in the redshift bins and to lower densities in the luminosity bins.
Comparing the peak-centred versions of the parameters to the fit-centred ones, we do not see a clear differences, although the normal distribution is favoured in more of the luminosity bins for the peak centred values.

The inner density slope, or cuspiness, parameters strongly disfavour normal models and prefer interpolated or skewed models.
For the fit-centred versions of the parameters, $\alpha$ and $\alpha_{50}$, the interpolated models show a bimodal structure with peaks at similar positions in all the luminosity and redshift bins.
This bimodal structure disappears when using the peak centred versions of the clusters,  $\alpha^*$ and $\alpha_{50}^*$, suggesting it could be induced by a population of clusters where the X-ray peak is not at the fit position.

If we models the distributions of the parameters assuming that they are the same for the whole sample, the evidence for non-normal distributions is significantly stronger.
This is seen for the central density and slope parameters, the scaled concentration, and even for the ellipticity.
However, evolution of parameters could also cause these non-normal distributions.

Therefore we do find evidence for significant non-Gaussian nor non-skew-normal distributions in some of our parameters, in particular the central density and inner slope.
The exceptions are the concentration and some of the 2D shape parameters ($\epsilon$ and $M_1$ to $M_4$).
The preferred model, however, will be dependent on the data quality and we see stronger evidence for non-normal distributions by examining the whole cluster sample.
Deeper eROSITA surveys will contain many more objects with a higher data quality than the current dataset.

When forward modelling in a cosmological analysis, the selection function should account for morphological variation.
In \cite{Clerc24}, the standard selection function marginalises over morphology, based on an input distribution of cluster profiles, although there is a version of the selection parametrised on $EM_0$, a concentration-like quantity.
If there were significant non-normal distributions of morphological parameters not present in the input profiles, this could potentially bias the selection model.
The most important morphological parameter for selection is the concentration, where we find normal distribution models are reasonable in individual redshift and luminosity bins, although the combined distribution prefers more complexity.
In addition, the redshift/luminosity evolution of concentration is relatively small or moderate.
The near normal concentration distribution is helpful for modelling the selection function.
To test this further, one could also include a parametrisation for concentration in the selection function and the cosmology forward model.

\section{Conclusions}
\label{sect:concl}
We investigated the intrinsic distributions and evolution of some of the morphological parameters of clusters detected in the eROSITA eRASS1 sky survey using a Bayesian framework, taking account of the selection of clusters in the survey.
Using this technique and assuming that the parameters are described by a scaling relation with redshift and luminosity and are distributed normally, we constrained the parameters of their evolution.

The concentration measured in a scaled aperture evolves positively with luminosity, as expected if cool core clusters are more luminous than non-cool core systems, and negatively with redshift, implying that low redshift clusters are more concentrated.
The scaled central density evolves similarly with luminosity with concentration, but does not significantly evolve in redshift.
The density at fixed physical radius, has a much stronger evolution with luminosity.
The concentration in a fixed aperture has a reduced evolution with luminosity, although its evolution depends on how the centre of the cluster is chosen.
If using the cluster peak as the cluster centre, its luminosity evolution is consistent with zero.
The negative luminosity and positive redshift evolution of the central slope is more difficult to understand.
However, this could be affected by the complex distributions of the parameters found within bins of redshift or luminosity, by observational measurement processes, or could be a real astrophysical effect.
The 2D shapes of the clusters are consistent with no evolution with luminosity or redshift.

We investigated the intrinsic distribution of parameters as a function of luminosity and redshift, fitting normal, skew normal, dual normal, and interpolated models.
We do not find a particular distribution is preferred in all bins for parameters.
The exceptions are the concentration and ellipticity, which favour a normal distribution in most cases.
The central density parameter distribution favours interpolated or skew models in different bins and often strongly disfavours a normal distribution, implying a more complex model is required.
The central slope or cuspiness parameter usually favours skew normal or interpolated models.
Fitting models the whole cluster population, shows stronger evidence for non-normal distributions in many cases, in particular the central density, $c_{500}$ and cuspiness.

The intrinsic distribution of parameters is compared against that for SZ selected cluster samples, which should not be biased by cool cores.
Despite modelling using the selection function and looking at the difference in concentration for the same clusters, we find that eROSITA clusters remain more concentrated than those in SZ surveys, with around 15 to 35\% more flux within $0.1R_{500}$.
The differences could be due to selection effects on the SZ detected objects, problems with our selection function or there are observation effects which might bias the concentration in eRASS1 detected clusters.
In addition, the eRASS1 clusters show central density gradients which are considerably larger than the SZ objects.
Despite the difference in concentration and inner slopes, we find little difference in the scaled central densities for the clusters between the eRASS1 and SZ selected objects.
In addition, the 2D shape of objects measured by ellipticity, slosh and multipole magnitudes looks rather similar between the eRASS1 and SPT samples, although Planck-selected clusters are less elliptical.

This analysis leads the way for the study of the intrinsic morphology of still-larger samples of clusters detected by eROSITA.
Deeper eROSITA surveys will also reduce the statistical uncertainty for many of the parameters for the brightest clusters.
It will also be important to make more detailed studies of the selection function to study the intrinsic distribution of these parameters.

\begin{acknowledgements}
This work is based on data from eROSITA, the soft X-ray instrument aboard SRG, a joint Russian-German science mission supported by the Russian Space Agency (Roskosmos), in the interests of the Russian Academy of Sciences represented by its Space Research Institute (IKI), and the Deutsches Zentrum für Luft- und Raumfahrt (DLR). The SRG spacecraft was built by Lavochkin Association (NPOL) and its subcontractors, and is operated by NPOL with support from the Max Planck Institute for Extraterrestrial Physics (MPE).

The development and construction of the eROSITA X-ray instrument was led by MPE, with contributions from the Dr. Karl Remeis Observatory Bamberg \& ECAP (FAU Erlangen-Nuernberg), the University of Hamburg Observatory, the Leibniz Institute for Astrophysics Potsdam (AIP), and the Institute for Astronomy and Astrophysics of the University of Tübingen, with the support of DLR and the Max Planck Society. The Argelander Institute for Astronomy of the University of Bonn and the Ludwig Maximilians Universität Munich also participated in the science preparation for eROSITA.

The eROSITA data shown here were processed using the eSASS/NRTA software system developed by the German eROSITA consortium.

E.B., V.G., A.L. and X.Z. acknowledge financial support from the European Research Council (ERC) Consolidator Grant under the European Union’s Horizon 2020 research and innovation program (grant agreement CoG DarkQuest No 101002585).
V.G. acknowledges the financial contribution from the contracts Prin-MUR 2022 supported by Next Generation EU (M4.C2.1.1, n.20227RNLY3 The concordance cosmological model: stress-tests with galaxy clusters).
A.L. acknowledges the supports from the National Natural Science Foundation of China (Grant No. 12588202). A.L. is supported by the China Manned Space Program with grant no. CMS-CSST-2025-A04.

\end{acknowledgements}

\bibliographystyle{aa}
\bibliography{refs}

\begin{appendix}
  \section{Simulations}
  \label{appen:sims}

To calculate our selection function as a function of morphological parameter, we made simulations of clusters and attempted to detect them.
The approach we used was much simplified over the simulations of \cite{Comparat20}.
Rather than include a proper clustered AGN background, we simulated a single cluster at a time, assuming a smooth point source free X-ray background map.
To simulate a cluster, we randomly chose positions within the sky regions used for eROSITA cluster detection.
The object was randomly assigned a log $L_{500}$ from a regularly spaced grid of 21 values between $42.5$ and $45.0$ log~erg~s$^{-1}$.
Redshifts were randomly taken from the list $0.02$, $0.035$, $0.05$, $0.07$, $0.1$, $0.15$, $0.2$, $0.3$, $0.4$, $0.5$, $0.6$, $0.7$, $0.8$, $1.0$ and $1.2$.
In total, around $6 \times 10^{5}$ objects were simulated.

The covariance matrix method described in \cite{Comparat20} was used to generate a random cluster mass, temperature and emissivity profiles.
Clusters were assigned to a grid point by generating a mass function \citep{Tinker08} for the relevant redshift and assigning randomly generated clusters nearby in mass and redshift.
Those clusters with luminosities close to the grid point value were assigned to it.
Rather than use the emissivity profile directly, which can be noisy  or may not be a valid 3D projection, we fit a full Vikhlinin density model \citep{Vikhlinin07} after projection to the emissivity profile, and use the best fitting model to generate a new emissivity profile for simulation.
Model concentration, density and cuspiness values were computed from this fitted model profile.

To create a simulated eROSITA event list, we used \texttt{sixte} 3.1.1 \citep{Dauser19}, providing an eRASS1 input attitude file.
The input file describing the model for \texttt{sixte} contained an image of the cluster following the emissivity profile with a symmetric shape and a model spectrum given by the cluster temperature and Galactic absorption for the sky position.
The cluster source was defined to have a flux such that the X-ray luminosity within $r_{500}$ matched the chosen luminosity.
No AGNs were included in the simulation.
An X-ray background model was also included, taken from wavelet-filtered sky maps of the real eRASS1 sky, where structures below scales of around 30 arcmin were removed.
The simulated flux of these maps was scaled to produce the count rate within our 0.2-2.3 keV band, once the \texttt{sixte} particle background model rate was subtracted.

The cluster region was simulated using a box of dimensions $8 r_{500}$, with a minimum size of 1.5 deg. 
The output event files from \texttt{sixte} were merged into a single event file and then filtered using the standard eROSITA detector mask, flags, patterns and the good time intervals (GTIs) used for eRASS1.
Images in the 0.2-2.3 keV band were created.
Exposure maps were taken from the eRASS1 all sky exposure maps.

To replicate cluster detection, we used the \texttt{eSASS} \citep{Brunner22} version \texttt{eSASSusers\_240410\_0\_3} on the simulated image.
\texttt{erbox} was used on the simulated image to make a list of sources (with \texttt{likemin=6}, \texttt{nruns=2}, \texttt{boxsize=4} and \texttt{bkima\_flag=N}).
This source list was supplied to \texttt{erbackmap} to make an initial background map (using \texttt{scut=0.00005}, \texttt{mlmin=6}, \texttt{maxcut=0.5}, \texttt{smoothval=15}, \texttt{snr=40}, \texttt{smoothmax=360}).
\texttt{erbox} was used a second time using the background map (with \texttt{likemin=4} and \texttt{bkima\_flag=Y}).
\texttt{erbackmap} was run a second time to produce a new background map based on the generated source list.
We ran \texttt{erbox} again using the previous background map and made a new background map using \texttt{erbackmap}.

As the high luminosity, nearby clusters produce a large number of events in their output files, the photon detection mode of the \texttt{ermldet} maximum likelihood detection software is unusably slow for these clusters.
Therefore we ran it up to twice, once in image mode, and if the cluster was not detected with a high enough significance, in photon mode.
The threshold we use to decide to use photon mode is if $\mathcal{L}_\mathrm{ext}<30$, well above the threshold of $6$ used to select the cosmology sample.
\texttt{ermldet} was run with  parameters \texttt{likemin=5}, \texttt{extlikemin=3}, \texttt{cutrad=15}, \texttt{multrad=20}, \texttt{extmin=2}, \texttt{extmax=15}, \texttt{nmaxfit=4} and \texttt{nmulsou=2}.
For the image mode we set \texttt{shapelet\_flag=no} and \texttt{photon\_flag=no}, while for photon mode these are both set to \texttt{yes}.

A cluster is detected if there is an extended object within radii of $6$ ($z \le 0.05$), $4$ ($0.05 \le z<0.2$), $3$ ($0.2 \le z < 0.4$) and $2$~arcmin ($z \ge 0.4$) from the input cluster position.
For those clusters we detect, we run \texttt{MBProj2D} on the images (excluding TMs 5 and 7), to obtain the total number of cluster counts, the cluster luminosity, concentration, central density and cuspiness.
The MBProj2D modelling is described in Paper I.

There are some limitations on the accuracy of the simulations from which the selection function have been derived, some of which we discussed in Paper I.
Firstly, the simulations are of spherical clusters with no contaminating point sources.
To calculate the parameters which depend on cluster radius, we assume that the input cluster mass is correct, rather than obtaining it from the simulated data.
We also do not account for redshift uncertainties for the objects, although this should be a relatively small effect given the quality of the X-ray detections.
Another potential shortcoming is that clusters simulated individual with random rather than correlated positions, underestimating effects due to nearby objects see in real systems \citep[e.g.][]{Ramos19,Spinelli25}.

\section{Modelling the selection function}
\label{appen:seln}
The simulation and detection procedure above describes the standard eROSITA detection pipeline.
The clusters we study also have additional selection applied of $\mathcal{L}_\mathrm{ext}>6$ and a minimum of 50 or 100 counts.
These selections are applied after the detection is run on the simulated clusters, to further restrict those clusters which are detected.

After studying the results of the simulations, we found that the cluster selection can be well described by a function of three values, the redshift, the morphological parameter in question, and a log count-like quantity, $Q$.
We define $Q$ as
\begin{multline}
  Q(L_{500},z,t,N_\mathrm{H}) = \log L_{500} -  2 \log \left[ D_L(z)/D_L(z=0.2) \right] + \\
  \log(t / t_\mathrm{ref}) + A(L_{500}, z, N_\mathrm{H}),
\end{multline}
where $L_{500}$ is the 0.2-2.3 keV cluster luminosity, $D_L(z)$ is the luminosity distance of an object at the given redshift, $t$ is the exposure time, $t_\mathrm{ref}$ is a reference exposure time ($90.8$s), and $A$ accounts for photoelectric absorption by our Galaxy.
$A$ is the median log difference in flux between an absorbed (with column density $N_\mathrm{H}$) and unabsorbed spectral model, with temperatures sampled from the simulated clusters with the given redshift and luminosity.
For a particular cluster, the only part of $Q$ which varies during the modelling analysis is $L_{500}$.
We did not find a significant effect on the selection function due to the background count rate, nor the exposure time once it is included in $Q$.

We fit a model to the simulated clusters in this three-dimensional space.
The model is a sum of Gaussian components, each with a free central position, normalisation and covariance matrix.
The selection function must lie in the range $[0,1]$ for any set of parameters.
Therefore we apply the logistic sigmoid function to the sum to ensure the output is in this range.
\begin{equation}
  I(\mathbf{x}) = \frac{1}{ 1 + \exp \left( - \sum_{i=1}^{N} A_i \, G_i(\mathbf{x}) \right) },
\end{equation}
where $G_i$ is a multivariate Gaussian function with mean $\boldsymbol\mu_i$ and covariance matrix $\boldsymbol\Sigma_i$, $N$ is the number of components and $A_i$ is the normalisation of component $i$ (which can be negative). $\mathbf{x}$ is the three-dimensional vector $(Q, \log z, Y)$, where $Y$ is the morphological parameter in question.

This function is fitted to the results of the simulations binned in the three-dimensional space.
$Q$ is binned into 80 equal-sized bins between $40.2$ and $47.0$.
We use 20 bins between $-2.0$ and $0.0$ for concentration parameters.
For $n_\mathrm{s}$ we use 80 bins between $-2.0$ and $2.5$.
For $n_{50}$ we use 80 bins between $-5.0$ and $0.0$.
For cuspiness, we use 80 bins between $-3$ and $3$.
The redshift values use the same grid as in Appendix \ref{appen:sims}.

In each bin we know the total number of simulations done and the number which resulted in a detection.
The binomial likelihood for $n$ trials with $k$ detections can be calculated.
To fit the selection function to the simulation data we maximise the total likelihood of all the bins.
It is difficult to choose the best value of $N$ for the analysis.
We tested different values, finding that $N=8$ was a reasonable compromise between modelling the shape of the function and being able to find a likelihood maximum in a reasonable time.
To prevent the model becoming undefined beyond the range of the simulations, for each of the redshift grid points, we added fake simulations in alternate bins with extreme values of $Q$ where no simulations were done.
For $Q$ values lower than the $0.1$ percentile for that redshift the $k$ was set to $0$, while $k$ was set to $n$ above the $99.9$ percentile, with $n$ set to $6$.

\begin{figure*}
  \includegraphics[width=\textwidth]{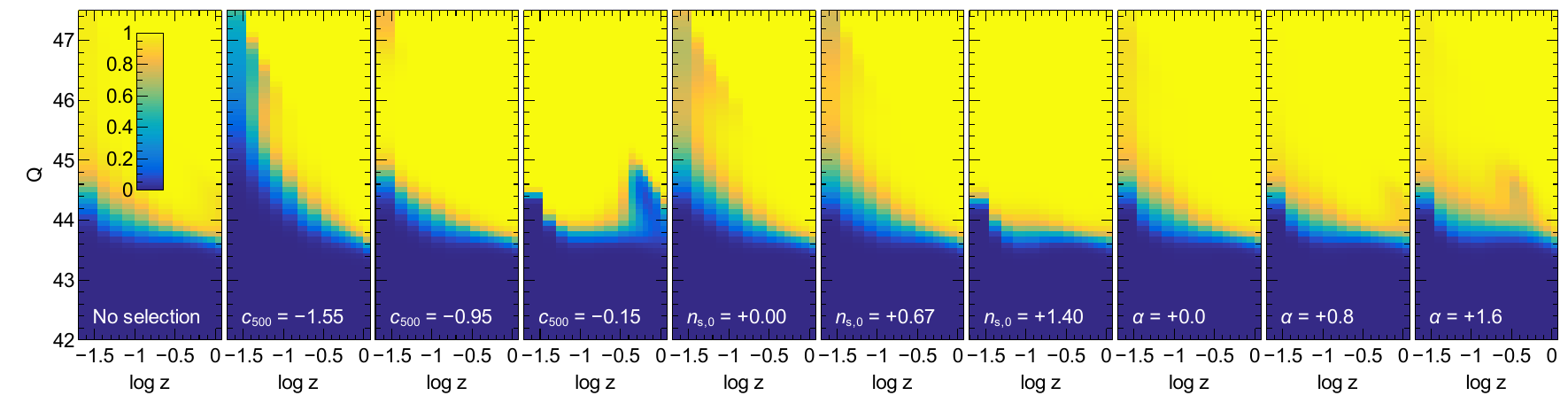}
  \caption{Selection function for different morphological selections.
    The selection function model for the 50 count subsample is shown as a function of $Q$ and $\log z$.
    From left to right, are shown no morphological selection, and selection for low, medium and high values of $c_{500}$, $n_{\mathrm{s},0}$, and $\alpha$.
  }
  \label{fig:selfn}
\end{figure*}

We found in Paper I that the selection function is not strongly affected by the parameters which affect the 2D shape of the cluster, rather than its profile, i.e. $\epsilon$, $H$ and $M_1$ to $M_4$.
In this case for these parameters, the selection function does not include the parameter and is only modelled in two dimensions.
Figure~\ref{fig:selfn} shows the selection function as a function of $Q$ and $\log z$ for different morphological parameter values.
Although the selection function is seen to vary relatively strongly at high redshift with concentration, it is harder to see the effect of the increase in central density, because higher values ($\sim 1.9$) are necessary to see the effect here.

We compared our selection function with the standard eROSITA one \citep{Clerc24} to check for consistency.
Taking an exposure time of 100s and $N_\mathrm{H}=3\times 10^{20}$~cm$^{-2}$, we compared the results for a grid of redshift and luminosity, converting the input to count rates as input for the standard selection function.
We find that the there is reasonable agreement, except in the lowest 0.05 redshift bin.
In general, the selection function generated in this paper is a little steeper, and reaches a value of 0.5 at most 0.1 dex lower in luminosity than the standard selection function.
A likely reason for the differences between our selection function and the standard eROSITA ones, is that we include neither clustered background point sources nor other extended sources.

\section{Scaling relation parameters}
\label{appen:scaling}

\begin{figure*}
  \centering
  \includegraphics[width=\columnwidth]{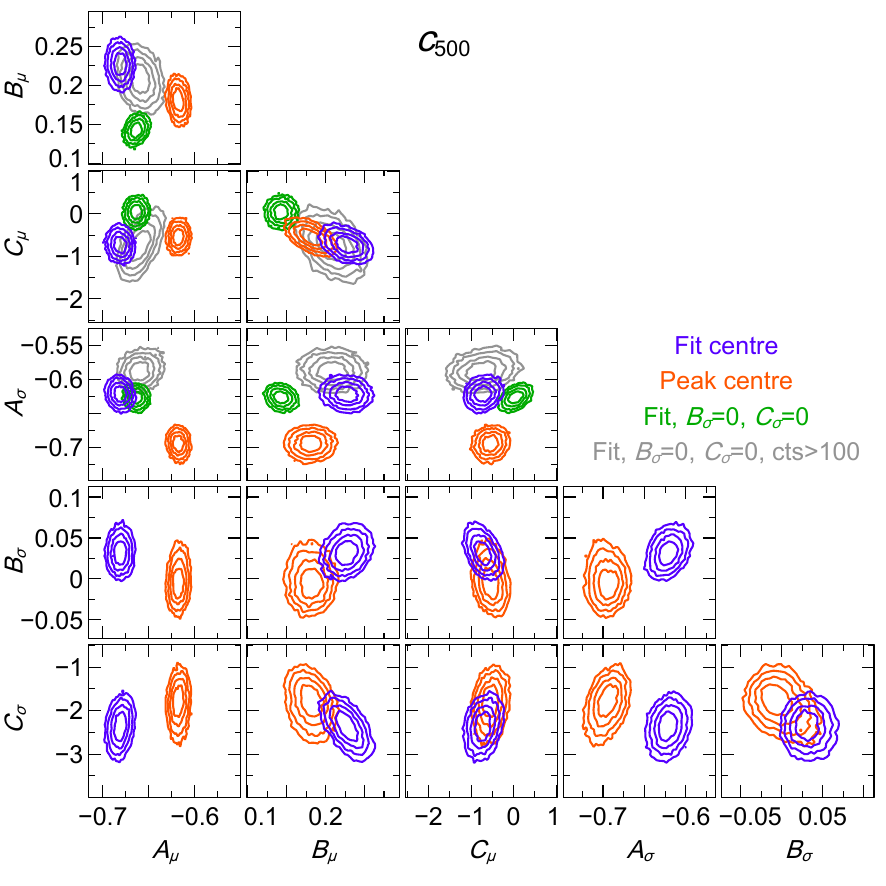}
  \includegraphics[width=\columnwidth]{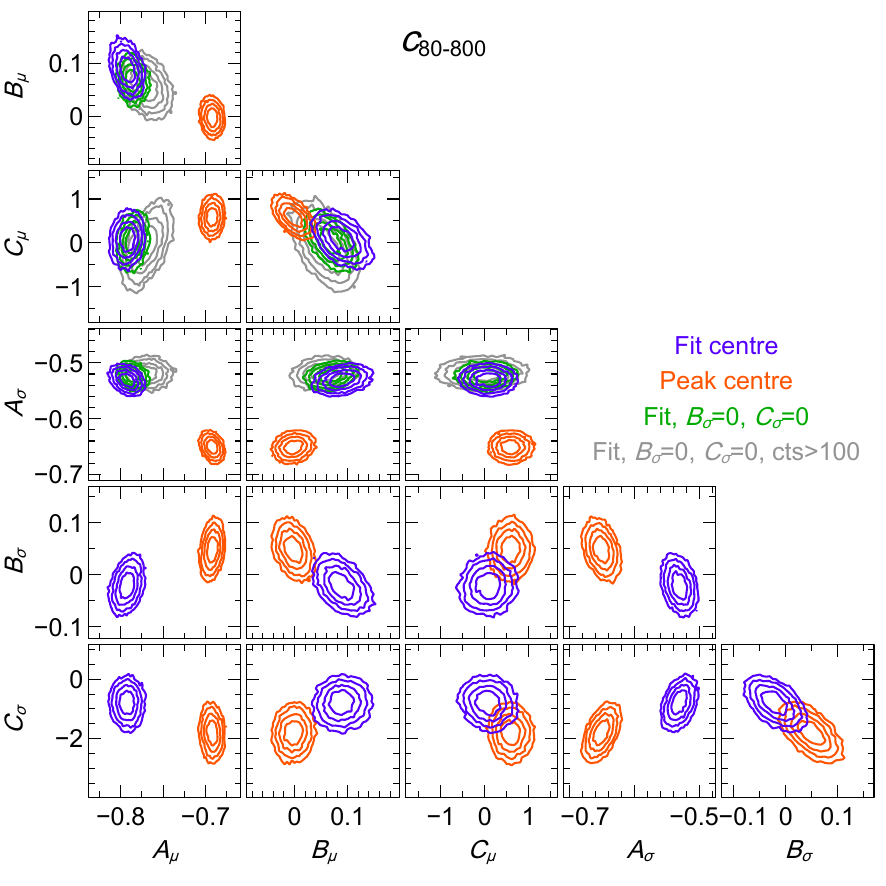}
  \includegraphics[width=\columnwidth]{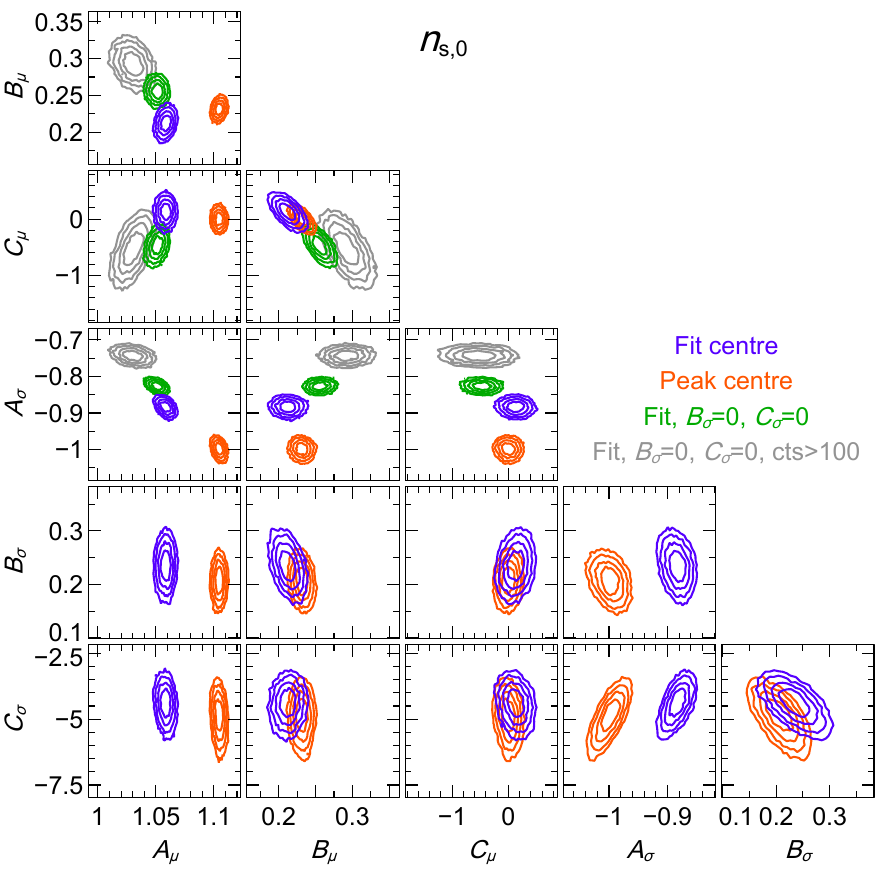}
  \includegraphics[width=\columnwidth]{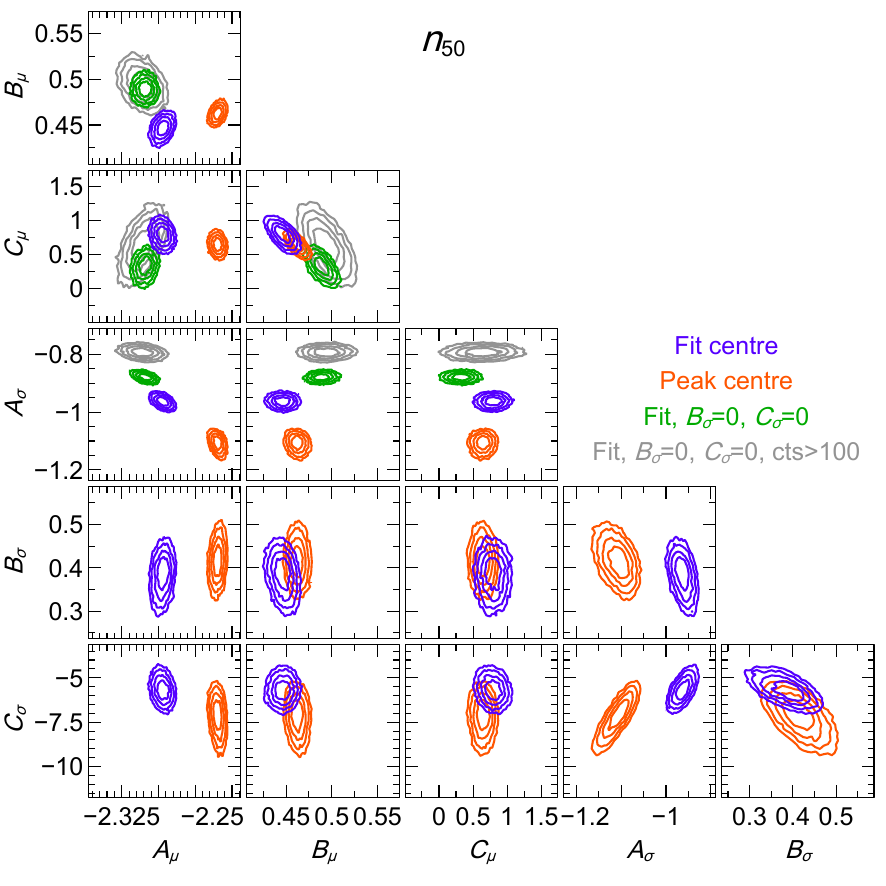}
  \caption{
    MCMC corner plots for the analysis of the scaling relation of concentration, $c_{500}$ (top left) and $c_{80-800}$ (top right), and density, $n_{\mathrm{s},0}$ (bottom left) and $n_{50}$ (bottom right).
    The contours contain 39.3, 67.5, 86.4 and 95\% of the samples.
    The contours are shown for the analysis with evolution in mean and width, where the centre positions of the clusters are fitted for, the same but using the peak cluster positions, a scaling relation with no evolution in width, with the fitted centres, and the same but only using clusters with more than 100 counts instead of 50 counts.
  }
  \label{fig:corner_conc_ne}
\end{figure*}

\begin{figure*}
  \centering
  \includegraphics[width=\columnwidth]{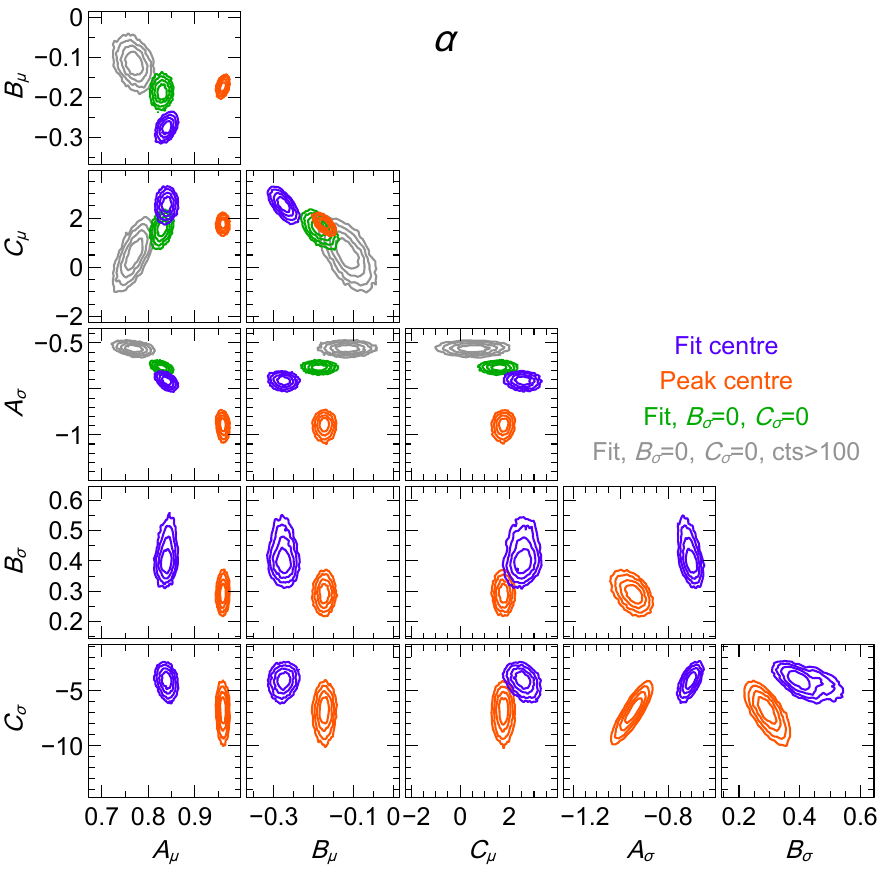}
  \includegraphics[width=\columnwidth]{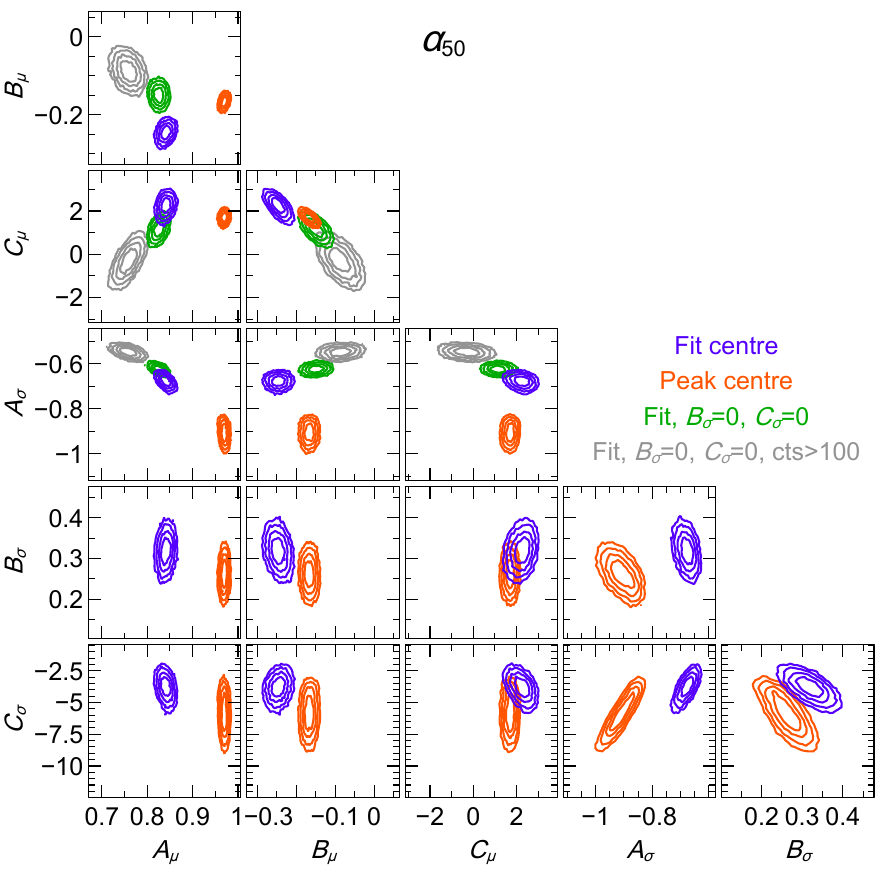}
  \includegraphics[width=\columnwidth]{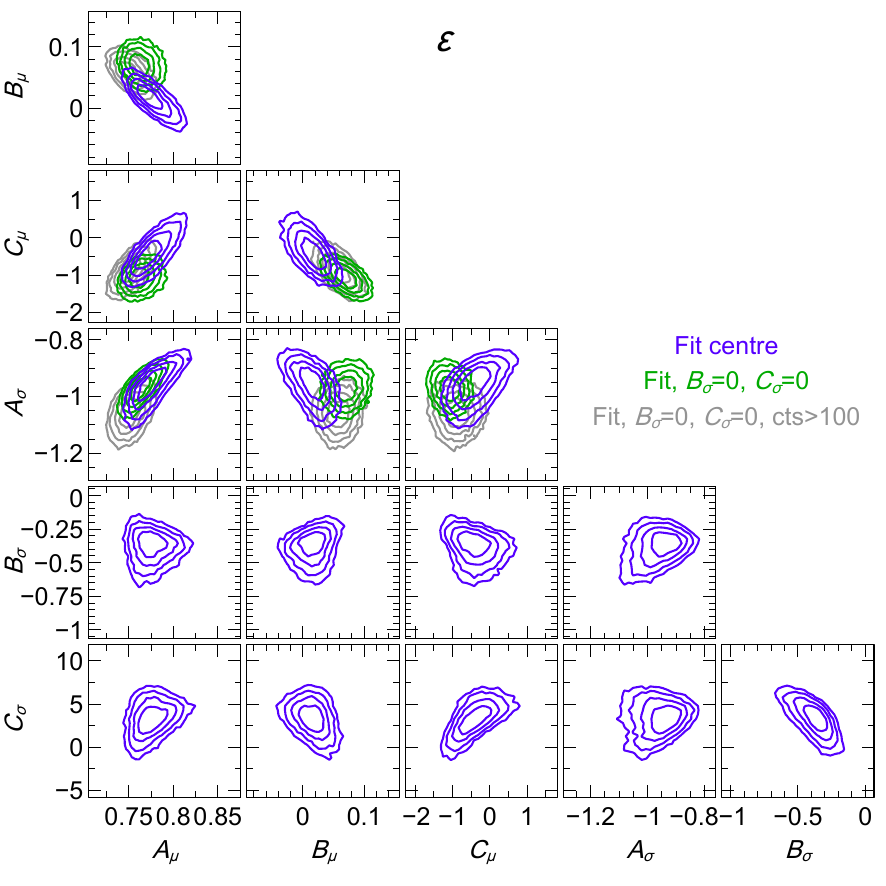}
  \includegraphics[width=\columnwidth]{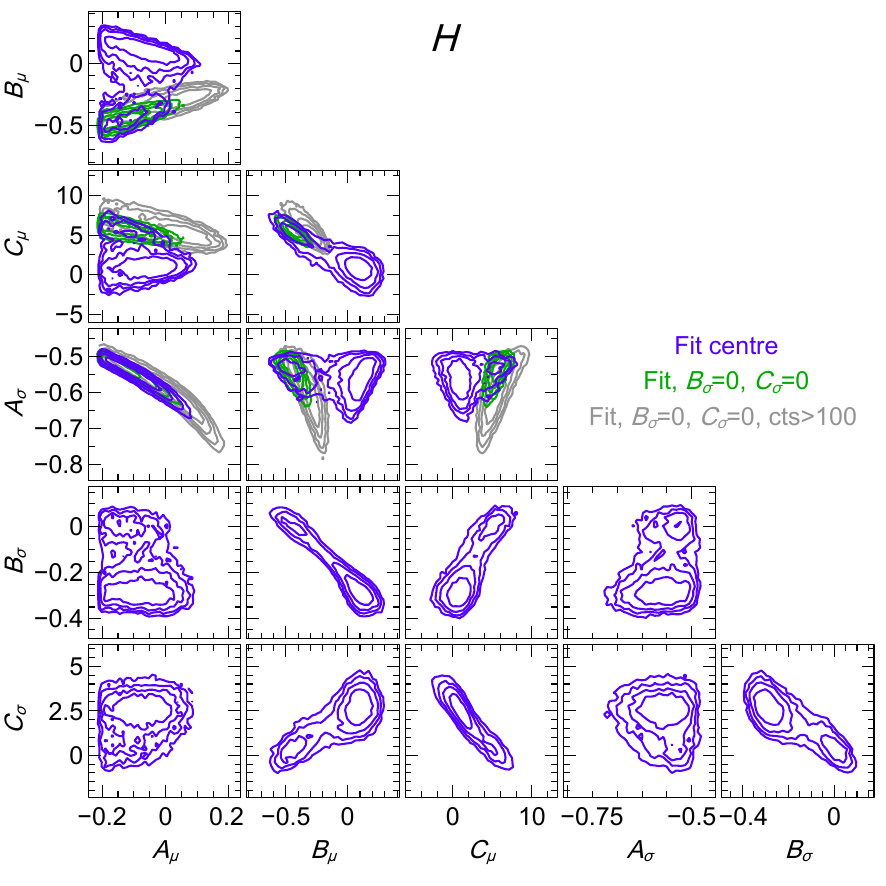}
  \caption{
    MCMC corner plots for the analysis of the evolution of inner density slope, $\alpha$ (top left), $\alpha_{50}$ (top right), ellipticity, $\epsilon$ (bottom left) and slosh, $H$ (bottom right).
    The description is the same as Fig.~\ref{fig:corner_conc_ne}, except that cluster position is always fitted for ellipticity and slosh.
  }
  \label{fig:corner_alpha_el_sl}
\end{figure*}

Table \ref{tab:conc_scaling} lists the median values from the MCMC chain and their $1\sigma$ posterior probability widths for the two different cluster samples.
Corner plots showing the distribution of parameter values in the MCMC chains are shown in Fig.~\ref{fig:corner_conc_ne} and \ref{fig:corner_alpha_el_sl}.
In these corner plots we show the results for the standard analysis (clusters with more than 50 counts, allowing an evolution of the distribution width and fitting for the cluster position), a second where the cluster is fixed at the peak position, a third where the width evolution parameters are fixed at 0, and finally with no evolution in width but only studying clusters with more than 100 counts.

\begin{table*}
  \caption{Parameters and their one-dimensional uncertainties for their luminosity scaling relations assuming normal scatter.}
  \centering
  \small
  \begin{tabular}{cccccccc}
\hline
Relation & Min. cts. & $A_\mu$ & $B_\mu$ & $C_\mu$ & $A_\sigma$ & $B_\sigma$ & $C_\sigma$ \\ \hline
$c_{500}$ & $50$ & $-0.6812 \pm 0.0068$ & $0.227 \pm 0.014$ & $-0.72 \pm 0.19$ & $-0.622 \pm 0.011$ & $0.032 \pm 0.014$ & $-2.36 \pm 0.32$ \\
$c_{500}$ & $50$ & $-0.6637 \pm 0.0062$ & $0.1426 \pm 0.0092$ & $0.05 \pm 0.17$ & $-0.6266 \pm 0.0088$ & $0$ & $0$ \\
$c_{500}$ & $100$ & $-0.660 \pm 0.011$ & $0.208 \pm 0.019$ & $-0.73 \pm 0.36$ & $-0.588 \pm 0.014$ & $0$ & $0$ \\
$c_{500}^*$ & $50$ & $-0.6179 \pm 0.0057$ & $0.181 \pm 0.014$ & $-0.54 \pm 0.19$ & $-0.695 \pm 0.012$ & $-0.004 \pm 0.019$ & $-1.81 \pm 0.37$ \\
$c_{500}^*$ & $50$ & $-0.6103 \pm 0.0054$ & $0.1480 \pm 0.0097$ & $-0.28 \pm 0.15$ & $-0.6830 \pm 0.0093$ & $0$ & $0$ \\
$c_{500}^*$ & $100$ & $-0.6271 \pm 0.0096$ & $0.191 \pm 0.018$ & $-0.75 \pm 0.34$ & $-0.651 \pm 0.013$ & $0$ & $0$ \\
$c_{80-800}$ & $50$ & $-0.7924 \pm 0.0087$ & $0.089 \pm 0.023$ & $0.08 \pm 0.28$ & $-0.531 \pm 0.012$ & $-0.021 \pm 0.025$ & $-0.80 \pm 0.39$ \\
$c_{80-800}$ & $50$ & $-0.7860 \pm 0.0080$ & $0.068 \pm 0.020$ & $0.04 \pm 0.29$ & $-0.524 \pm 0.011$ & $0$ & $0$ \\
$c_{80-800}$ & $100$ & $-0.769 \pm 0.012$ & $0.057 \pm 0.026$ & $-0.05 \pm 0.42$ & $-0.519 \pm 0.013$ & $0$ & $0$ \\
$c_{80-800}^*$ & $50$ & $-0.6921 \pm 0.0063$ & $-0.002 \pm 0.017$ & $0.59 \pm 0.21$ & $-0.651 \pm 0.012$ & $0.049 \pm 0.025$ & $-1.80 \pm 0.42$ \\
$c_{80-800}^*$ & $50$ & $-0.6920 \pm 0.0065$ & $-0.002 \pm 0.016$ & $0.57 \pm 0.23$ & $-0.625 \pm 0.011$ & $0$ & $0$ \\
$c_{80-800}^*$ & $100$ & $-0.711 \pm 0.011$ & $0.034 \pm 0.021$ & $0.10 \pm 0.37$ & $-0.599 \pm 0.013$ & $0$ & $0$ \\
$\alpha$ & $50$ & $0.8394 \pm 0.0101$ & $-0.276 \pm 0.016$ & $2.55 \pm 0.31$ & $-0.706 \pm 0.023$ & $0.406^{+0.054}_{-0.037}$ & $-4.08 \pm 0.75$ \\
$\alpha$ & $50$ & $0.829 \pm 0.010$ & $-0.186 \pm 0.019$ & $1.57 \pm 0.32$ & $-0.634 \pm 0.016$ & $0$ & $0$ \\
$\alpha$ & $100$ & $0.769 \pm 0.019$ & $-0.115 \pm 0.029$ & $0.53 \pm 0.62$ & $-0.530 \pm 0.019$ & $0$ & $0$ \\
$\alpha^*$ & $50$ & $0.9603 \pm 0.0060$ & $-0.173 \pm 0.012$ & $1.75 \pm 0.19$ & $-0.947 \pm 0.036$ & $0.293 \pm 0.031$ & $-6.85 \pm 1.18$ \\
$\alpha^*$ & $50$ & $0.9575 \pm 0.0064$ & $-0.160 \pm 0.013$ & $1.57 \pm 0.22$ & $-0.827 \pm 0.015$ & $0$ & $0$ \\
$\alpha^*$ & $100$ & $0.894 \pm 0.013$ & $-0.099 \pm 0.020$ & $0.59 \pm 0.44$ & $-0.708 \pm 0.019$ & $0$ & $0$ \\
$\alpha_{50}$ & $50$ & $0.841 \pm 0.010$ & $-0.246 \pm 0.017$ & $2.24 \pm 0.34$ & $-0.678 \pm 0.021$ & $0.318 \pm 0.032$ & $-3.79 \pm 0.76$ \\
$\alpha_{50}$ & $50$ & $0.825 \pm 0.011$ & $-0.149 \pm 0.018$ & $1.15 \pm 0.33$ & $-0.624 \pm 0.015$ & $0$ & $0$ \\
$\alpha_{50}$ & $100$ & $0.759 \pm 0.018$ & $-0.088 \pm 0.026$ & $-0.29 \pm 0.56$ & $-0.548 \pm 0.018$ & $0$ & $0$ \\
$\alpha_{50}^*$ & $50$ & $0.9697 \pm 0.0060$ & $-0.167 \pm 0.012$ & $1.70 \pm 0.20$ & $-0.908 \pm 0.038$ & $0.261 \pm 0.033$ & $-5.86 \pm 1.34$ \\
$\alpha_{50}^*$ & $50$ & $0.9633 \pm 0.0065$ & $-0.152 \pm 0.012$ & $1.45 \pm 0.22$ & $-0.816 \pm 0.015$ & $0$ & $0$ \\
$\alpha_{50}^*$ & $100$ & $0.889 \pm 0.012$ & $-0.092 \pm 0.019$ & $0.07 \pm 0.41$ & $-0.714 \pm 0.018$ & $0$ & $0$ \\
$n_\mathrm{s,0}$ & $50$ & $1.0585 \pm 0.0043$ & $0.213 \pm 0.011$ & $0.13 \pm 0.15$ & $-0.883 \pm 0.014$ & $0.235 \pm 0.028$ & $-4.40 \pm 0.53$ \\
$n_\mathrm{s,0}$ & $50$ & $1.0511 \pm 0.0046$ & $0.2559 \pm 0.0097$ & $-0.47 \pm 0.16$ & $-0.827 \pm 0.010$ & $0$ & $0$ \\
$n_\mathrm{s,0}$ & $100$ & $1.0302 \pm 0.0082$ & $0.292 \pm 0.016$ & $-0.53 \pm 0.29$ & $-0.743 \pm 0.014$ & $0$ & $0$ \\
$n_\mathrm{s,0}^*$ & $50$ & $1.1046 \pm 0.0032$ & $0.2314 \pm 0.0078$ & $0.00 \pm 0.11$ & $-0.998 \pm 0.016$ & $0.207 \pm 0.025$ & $-4.88 \pm 0.62$ \\
$n_\mathrm{s,0}^*$ & $50$ & $1.0993 \pm 0.0035$ & $0.2473 \pm 0.0077$ & $-0.30 \pm 0.12$ & $-0.9287 \pm 0.0098$ & $0$ & $0$ \\
$n_\mathrm{s,0}^*$ & $100$ & $1.0731 \pm 0.0066$ & $0.291 \pm 0.012$ & $-0.50 \pm 0.24$ & $-0.846 \pm 0.012$ & $0$ & $0$ \\
$n_{50}$ & $50$ & $-2.2971 \pm 0.0038$ & $0.4465 \pm 0.0080$ & $0.79 \pm 0.11$ & $-0.963 \pm 0.015$ & $0.386 \pm 0.037$ & $-5.69 \pm 0.55$ \\
$n_{50}$ & $50$ & $-2.3091 \pm 0.0041$ & $0.4889 \pm 0.0079$ & $0.33 \pm 0.13$ & $-0.879 \pm 0.011$ & $0$ & $0$ \\
$n_{50}$ & $100$ & $-2.3110 \pm 0.0071$ & $0.494 \pm 0.013$ & $0.64 \pm 0.25$ & $-0.793 \pm 0.014$ & $0$ & $0$ \\
$n_{50}^*$ & $50$ & $-2.2598 \pm 0.0028$ & $0.4623 \pm 0.0062$ & $0.644 \pm 0.091$ & $-1.107 \pm 0.023$ & $0.415 \pm 0.036$ & $-7.13 \pm 0.85$ \\
$n_{50}^*$ & $50$ & $-2.2683 \pm 0.0029$ & $0.4809 \pm 0.0065$ & $0.445 \pm 0.100$ & $-0.987 \pm 0.011$ & $0$ & $0$ \\
$n_{50}^*$ & $100$ & $-2.2786 \pm 0.0057$ & $0.496 \pm 0.011$ & $0.54 \pm 0.21$ & $-0.885 \pm 0.013$ & $0$ & $0$ \\
$\epsilon$ & $50$ & $0.775 \pm 0.015$ & $0.016 \pm 0.020$ & $-0.41 \pm 0.40$ & $-0.956 \pm 0.056$ & $-0.371 \pm 0.096$ & $3.40 \pm 1.60$ \\
$\epsilon$ & $50$ & $0.763 \pm 0.011$ & $0.071 \pm 0.018$ & $-1.08 \pm 0.26$ & $-0.971 \pm 0.042$ & $0$ & $0$ \\
$\epsilon$ & $100$ & $0.752 \pm 0.012$ & $0.058 \pm 0.017$ & $-0.88 \pm 0.32$ & $-1.056 \pm 0.049$ & $0$ & $0$ \\
$H$ & $50$ & $-0.082 \pm 0.079$ & $0.05 \pm 0.29$ & $1.30^{+2.81}_{-1.46}$ & $-0.556 \pm 0.043$ & $-0.255^{+0.217}_{-0.063}$ & $2.28 \pm 1.27$ \\
$H$ & $50$ & $-0.119^{+0.075}_{-0.061}$ & $-0.441 \pm 0.062$ & $5.54 \pm 0.83$ & $-0.539 \pm 0.034$ & $0$ & $0$ \\
$H$ & $100$ & $0.030 \pm 0.097$ & $-0.299 \pm 0.077$ & $5.41^{+1.51}_{-1.19}$ & $-0.607 \pm 0.067$ & $0$ & $0$ \\
\hline
  \end{tabular}
  \tablefoot{
    The value and uncertainties are the median and $1\sigma$ ranges from the posterior probability distributions.
  }
  \label{tab:conc_scaling}
\end{table*}

\section{Distribution detailed results}
\label{appen:distn}

\begin{figure*}
  \centering
  \includegraphics[width=0.47\textwidth]{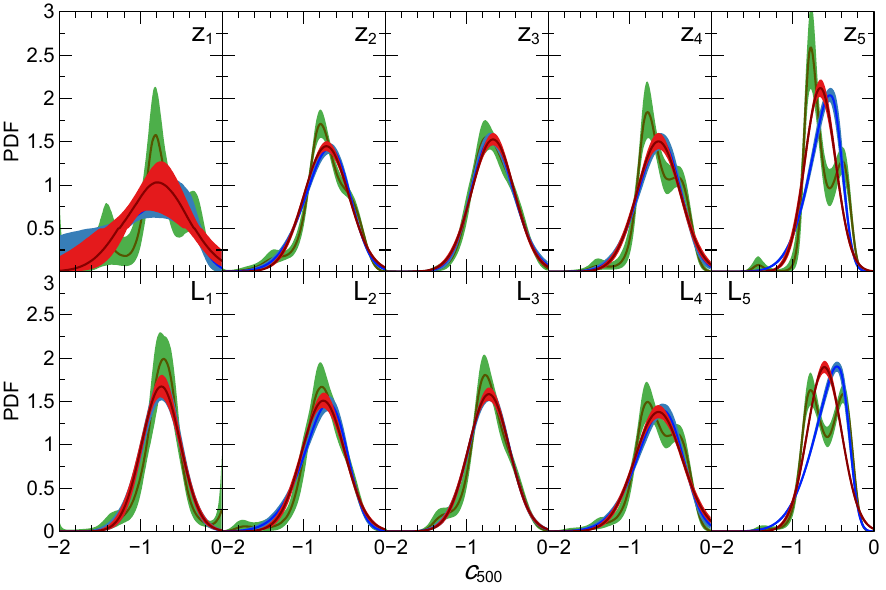}
  \includegraphics[width=0.47\textwidth]{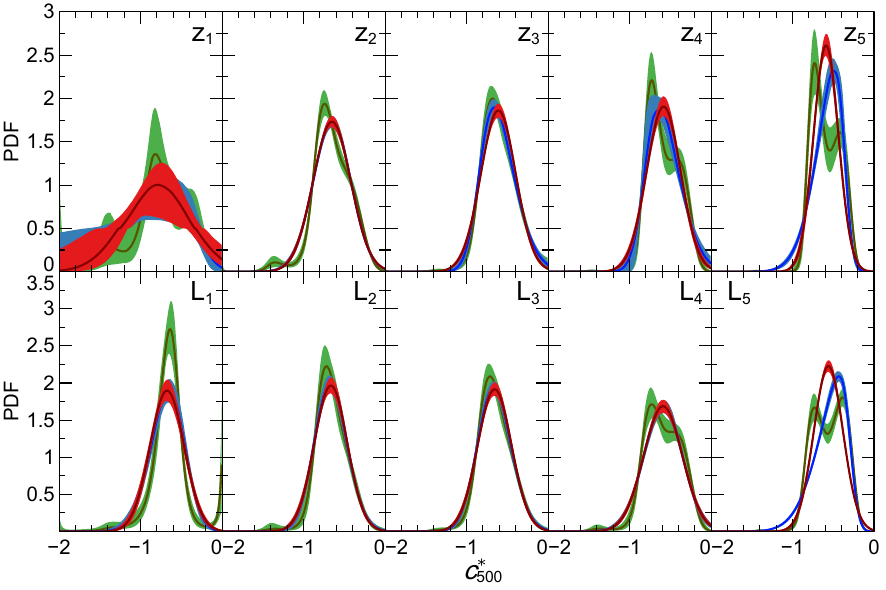}\\
  \includegraphics[width=0.47\textwidth]{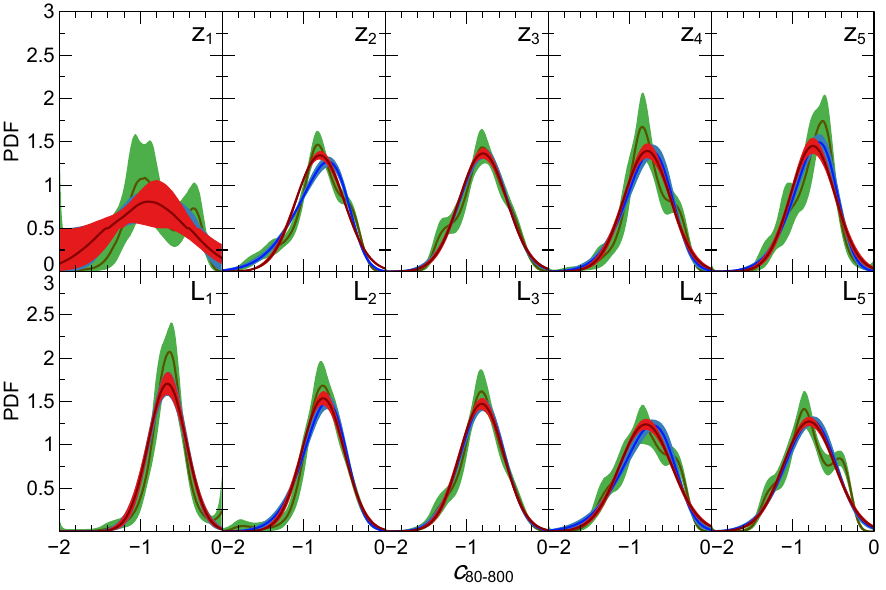}
  \includegraphics[width=0.47\textwidth]{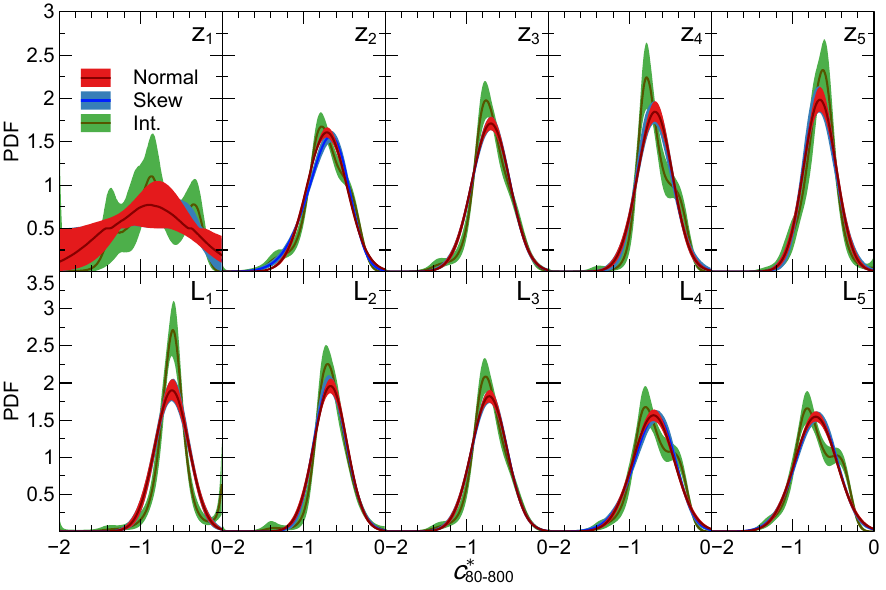} \\
  \includegraphics[width=0.47\textwidth]{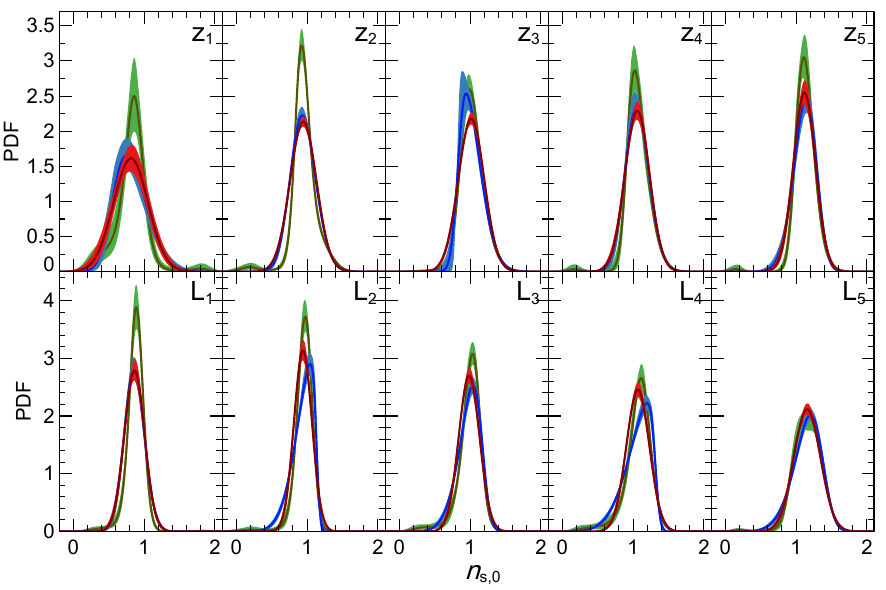}
  \includegraphics[width=0.47\textwidth]{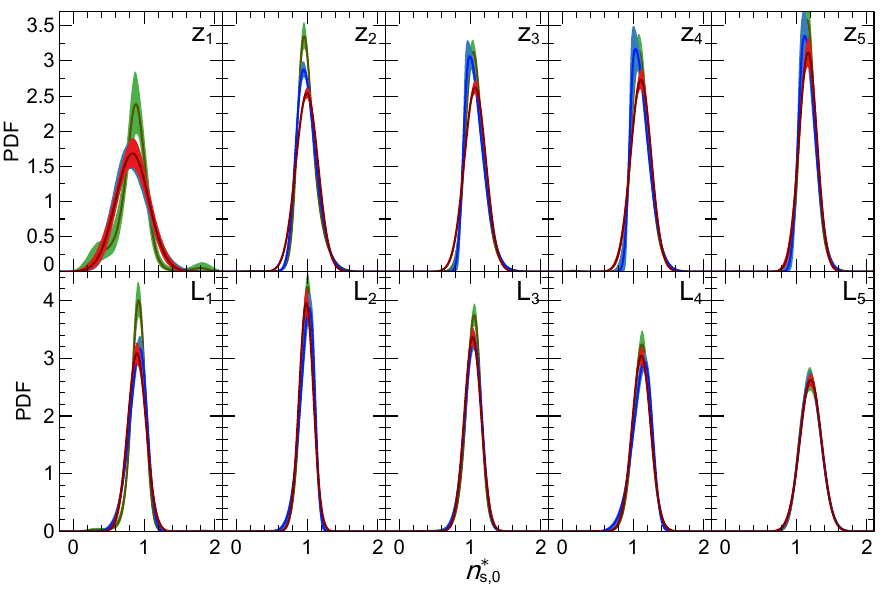}\\
  \includegraphics[width=0.47\textwidth]{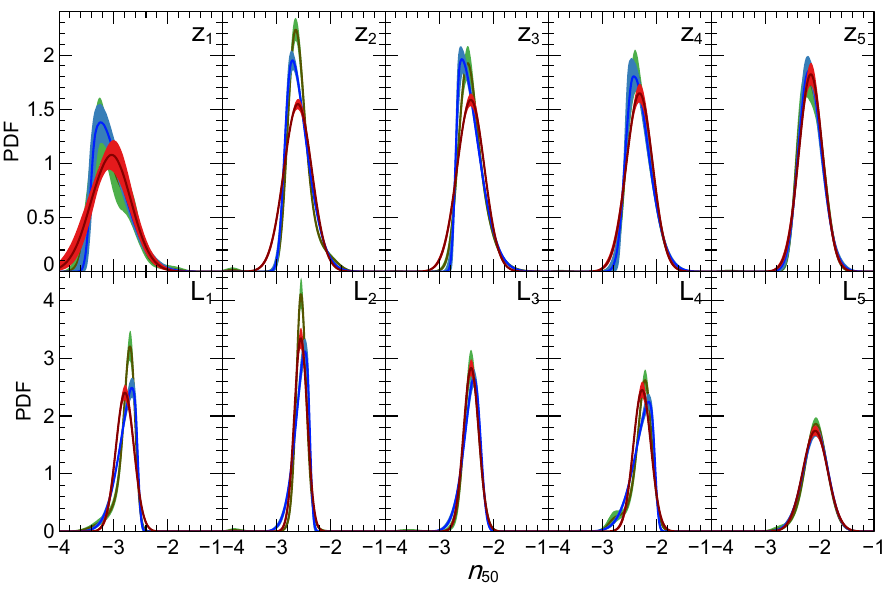}
  \includegraphics[width=0.47\textwidth]{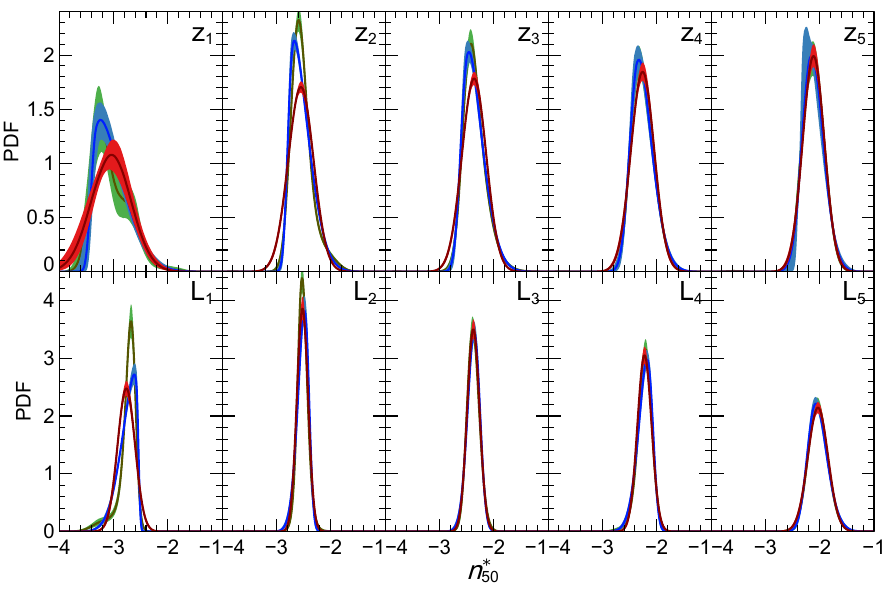}\\
  \caption{
    Distributions of concentration and density parameters in five bins of redshift and luminosity, after taking account of selection effects and the luminosity function.
    Shown are the distributions (median and $1\sigma$ range) for normal, skew normal and interpolated PDFs.
    The edges of the redshift bins are $0.0$, $0.1$, $0.2$, $0.3$, $0.4$ and $1.5$.
    The edges of the luminosity bins are $41.1$, $43.3$, $43.7$, $44.0$, $44.3$ and $45.6$ log\,erg\,s$^{-1}$.
  }
  \label{fig:dists_after_sel_1}
\end{figure*}

\begin{figure*}
  \centering
  \includegraphics[width=0.47\textwidth]{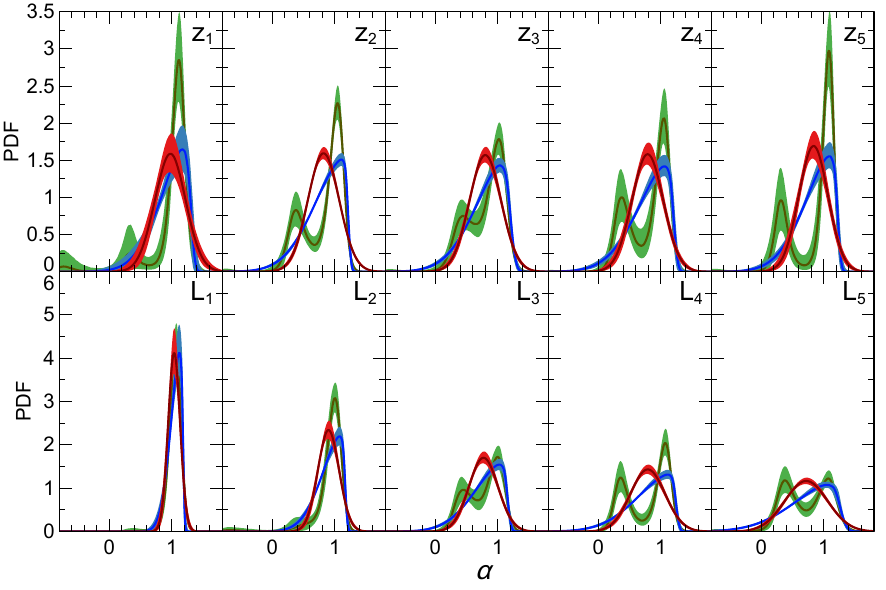}
  \includegraphics[width=0.47\textwidth]{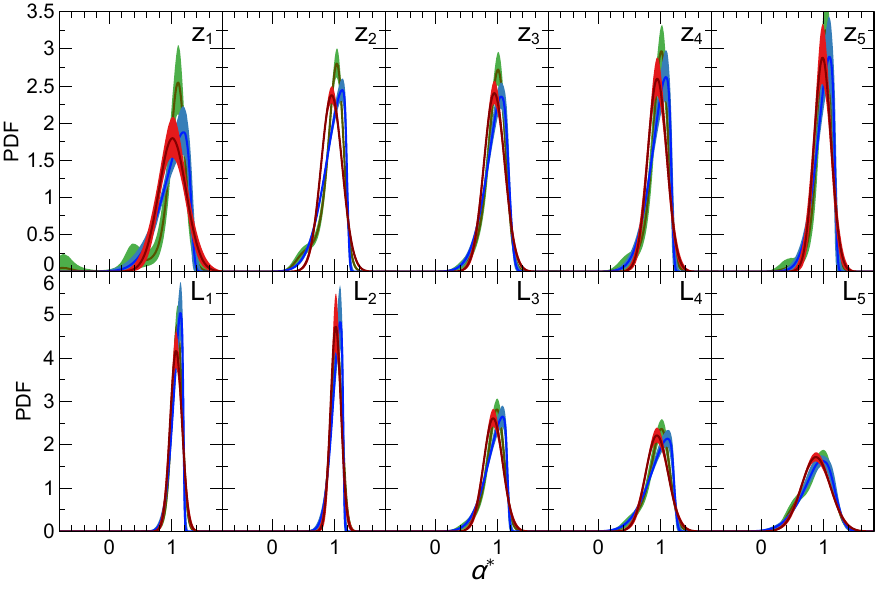}\\
  \includegraphics[width=0.47\textwidth]{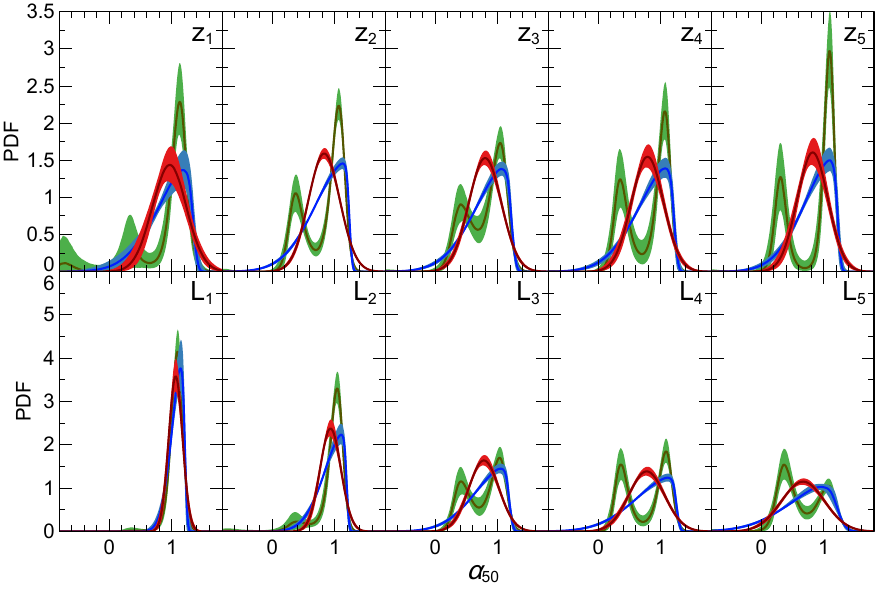}
  \includegraphics[width=0.47\textwidth]{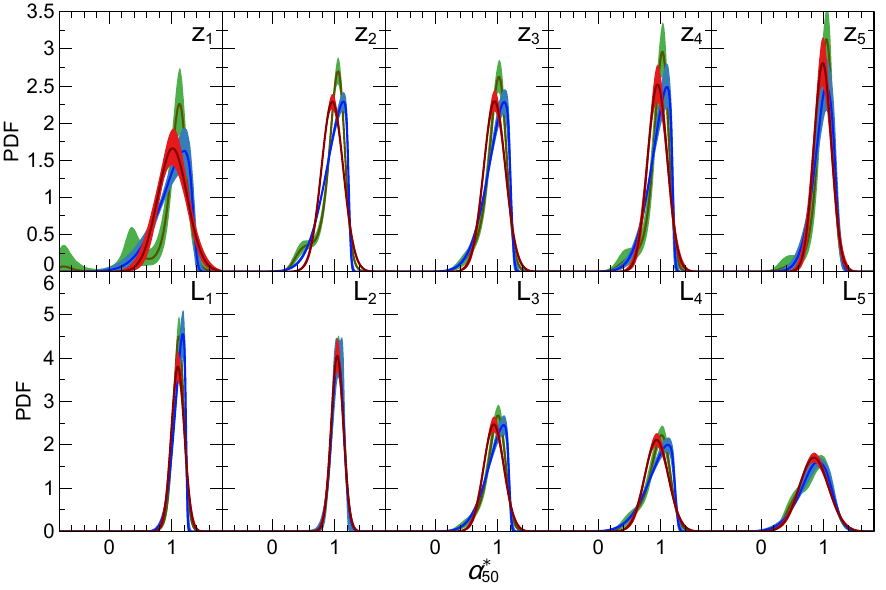}\\
  \includegraphics[width=0.47\textwidth]{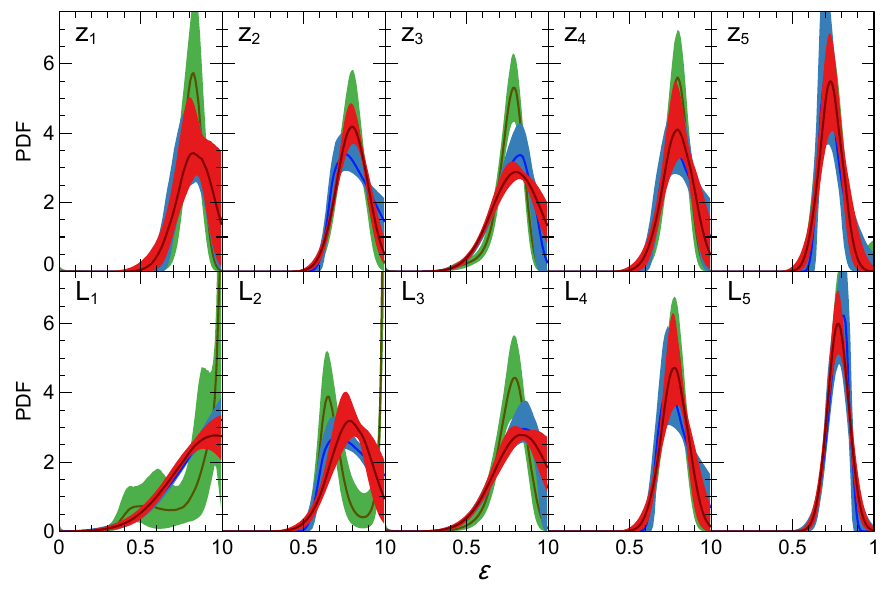}
  \includegraphics[width=0.47\textwidth]{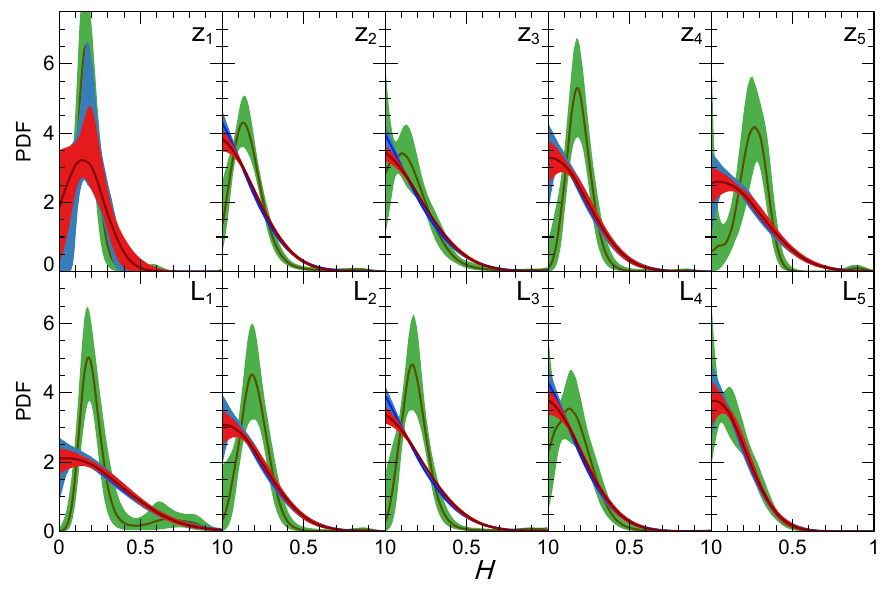}\\
  \caption{
    Distributions of cuspiness, ellipticity and slosh parameters, similarly to Fig.~\ref{fig:dists_after_sel_1}.
  }
  \label{fig:dists_after_sel_2}
\end{figure*}

Figure \ref{fig:dists_after_sel_1} and Fig.~\ref{fig:dists_after_sel_2} show distributions for three of the four models (normal, skew normal, and interpolated).
The parameters shown are the concentrations, the central densities, the cuspiness, the ellipticity and slosh.
The distributions were generated from the equal-weighted posterior samples produced by \texttt{UltraNest}.

\end{appendix}

\end{document}